\documentclass[aps,prc,preprint,groupedaddress,showpacs]{revtex4-1}

\usepackage{graphicx}
\usepackage{dcolumn}
\usepackage{color} 

\begin{document}


\title{Mass number and excitation energy dependence of the $\Theta _{eff}/\Theta _{rig}$ parameter of the spin cut-off factor in the
formation of an isomeric pair}


\author{S. Sud\'ar$^{1}$}
\email[Corresponding author. E-mail: ]{sudarsa@gmail.com}
\author{S.M. Qaim$^2$}
\affiliation{$^1$Institute of Experimental Physics, 
University of Debrecen, H-4010 Debrecen Pf. 105, Hungary.}
\affiliation{$^2$Institut f\"ur Neurowissenschaften und Medizin, INM-5: Nuklearchemie, Forschungszentrum 
J\"ulich GmbH, D-52425 J\"ulich, Germany}

\newcommand{\ones}{1.2}
\newcommand{\two}{1.2}

\date{\today}

\begin{abstract}
The $\Theta _{eff}/\Theta _{rig}$ parameter of the spin distribution of the level density of an excited nucleus was determined for 61 nuclei covering the mass range of 44 to 200. 
The experimental isomeric cross-section ratios for 25 isomeric pairs were compared with the model calculation to find the best fit to the experimental data. An isomeric cross-section ratio may be affected by more nuclei depending on the reaction. 
The model calculations were carried out with the TALYS code using the back-shifted Fermi gas model for the level density. The reduced $\chi ^2$ values were calculated to describe the deviation of the experimental data from the model calculation as a function of the $\eta =\Theta _{eff}/\Theta _{rig}$ parameter. 
The minimum of the reduced $\chi ^2$ curve defined the best probable $\Theta _{eff}/\Theta _{rig}$ parameter. An $\eta_d $ was introduced calculating the $\Theta _{eff}/\Theta _{rig}$ value from the low energy discrete levels of the nucleus. The $\eta /\eta_d $ values seem to be independent of the mass number, their average value near to one, but showing slight dependence on the odd-even characteristic of the proton and neutron number. The $\eta /\eta_d $ values also seem to be independent of the average excitation energy in the continuum, indicating  that  $\eta $ is independent of the excitation energy.
The mass number and (N-Z) dependence of the $\eta $ values were studied. The $\eta $ values for the nuclei with odd mass number show   a characteristic exponential decrease  as a function of the mass number or (N-Z). The $\eta $ values for the odd-odd type of nuclei appear to be constant, near one, up to the mass number 110 or (N-Z) of 13. 
Above these values, the $\eta $ values show characteristic exponential decrease like nuclei with the odd mass number.
The $\eta $ values for even-even type nuclei decrease exponentially up to A equal to 142 or (N-Z) equal to 20. The values are lower than those for the nearby odd-odd type nuclei. 
The $\eta $ values increase for nuclei with A between 142 and 156 and (N-Z) between 20 and 28, and they finally reach the value of odd-odd type systematics. This behavior indicates some individual properties of the $^{156}Gd$.
The odd-even effect of $\Theta _{eff}/\Theta _{rig}$ may change the calculated activation cross section significantly in some energy regions.
\end{abstract}

\pacs{21.10.Ma, 21.10.Hw, 23.35.+g, 24.10.-i, 24.60.Dr, 25.40.-h, 25.55.-e, 29.85.Fj}

\maketitle

\section{Introduction}

The spin distribution of the nuclear levels 
can be studied experimentally by investigating the rotational bands of the 
nuclei \cite{Mueller} or by studying the isomeric cross-section ratios.
The dependence of the isomeric ratio on the excitation energy of the product nucleus as well as on the 
spins of the two states concerned has been demonstrated by considering data for $(n,\gamma ), (n,2n), (n,p)$, $(n,\alpha )$ and $(\gamma ,n)$ reactions 
\cite{qaim72,qaim90,cserp,nesa2003,tatari15}.
In charged particle induced reactions, a comparison of measured isomeric cross-section ratios with values obtained from nuclear model calculations also sheds light on the spin distribution of levels involved 
\cite{DeYoung,qaim88,sudar96,stro97,Sudar2000}.

The level density distribution plays  a fundamental role 
 in the nuclear model calculations. 
Since Bethe's pioneering work \cite{Bethe} the nuclear level density problem has remained an active 
area of both theoretical and experimental studies. The original treatment was based on the 
non-interacting fermion gas in the nuclear volume, having equally spaced energy
 levels. Such a model corresponds 
to the zero-order approximation of a Fermi gas model.
The main parameter of the level density function, \textbf{a}, is determined mostly from the measured mean s-wave neutron resonance spacing. The low energy neutrons can excite only the levels of the compound nucleus having the spin of I=I$_t\pm$1/2, where I$_t$ is the spin of the ground state of the target nucleus. 
Therefore the level density parameter cannot be determined without knowing the spin distribution.

A common assumption in the global parametrization of the nuclear level 
density within the framework of the Fermi gas model is that the spin distribution 
is described by the formula
\begin{eqnarray}  \label{eq:spindep}
\frac{\rho_J}{\rho}=\frac{2J+1}{2 \sqrt{2\pi}\sigma^3}e^{-\frac{J(J+1)}{2\sigma^2}}
\end{eqnarray}
 where $\rho $ is the total level density, while $\rho_J$ is the
 density of spin-J levels without the $2J+1$ degeneracy factor. The parameter $\sigma $
 is known as the spin-cutoff parameter. The $\rho $, $\rho_J$ and $\sigma $ 
are all functions of the excitation energy.
 The spin-cutoff parameter can be expressed in terms of the expected value of the 
 $\roarrow J^2$ 
\begin{eqnarray} 
\sigma^2=\frac{1}{3}<\roarrow J^2>
\end{eqnarray}
In thermal ensembles it is common to define an effective moment of inertia 
 $\Theta _{eff}$ by the relation between $<\roarrow J^2>$ and temperature $T$, which we can write as
\begin{eqnarray} 
\Theta _{eff}=\frac{\hbar ^2}{T}\sigma^2
\end{eqnarray}
In the empirical parametrization, the $\sigma $ is determined using the rigid body moment
of inertia
\begin{eqnarray} 
\Theta _{rig}=\frac{2}{5}mA(r_0A^{1/3})^2,
\end{eqnarray}
where $r_0$ is the nuclear radius parameter, $A$ the mass number and $m$ is the nucleon mass. 
Introducing $\eta = \Theta _{eff}/\Theta _{rig}$, the square of the spin-cutoff 
parameter can be expressed as
\begin{eqnarray} 
\sigma^2=\eta \frac{\Theta _{rig}T}{\hbar ^2}
\end{eqnarray}

The temperature $T$ is related to the excitation energy of the nucleus $E_x$, back shift $\Delta$ and level density parameter $\textbf{a}$ as:
\begin{eqnarray} 
T=\sqrt{\frac{E_x-\Delta}{\textbf{a}}}
\end{eqnarray}
With this formula the above mentioned spin-cutoff parameter can be written as:
\begin{eqnarray} \label {eq:spincut}
\sigma^2=\eta \frac{\Theta _{rig}}{\hbar ^2}\sqrt{\frac{E_x-\Delta}{\textbf{a}}}
\end{eqnarray}
The full level density function is then described as
\begin{widetext}
\begin{eqnarray} \label{eq:totlvd}
\rho_J(E_x,J,\textbf{a},\Delta,\eta ,\Theta _{rig}(r_0,A))=\frac{2J+1}{2 \sqrt{2\pi}\sigma^3}e^{-\frac{J(J+1)}{2\sigma^2}}\frac{\sqrt{\pi}}{12}\frac{e^{2\sqrt{a(E_x-\Delta)}}}{\textbf{a}^{\frac{1}{4}}(E_x-\Delta)^{\frac{5}{4}}}
\end{eqnarray}
\end{widetext}

 The original Bethe formula was later refined
 to take into account the shell and pairing effects. One of the refined formulas is
 the back-shifted Fermi gas (BSFG) model \cite{dilg}. 
Recent theoretical studies are based on microscopic statistical model 
 \cite{Hilaire01,Demetriou} which is beginning to approach 
the quality of the semi-empirical models, but its 
results are available only in the form of numerical data.
Theoretical studies have shown that for high spin values the spin dependence of the level 
density can deviate from the simple formula given above 
 \cite{Paar}. Studies on the spin distribution are interesting
 both from the theoretical 
\cite{Quentin} and experimental points of view.
The theoretical analysis has also shown the excitation energy dependence 
of the spin-cutoff parameter \cite{Agrawal}.

When formula (\ref{eq:totlvd}) is used in the nuclear reaction model calculation the parameters ($\textbf{a},\Delta,\eta $) are usually fixed. Their values have to be determined from other independent experiments. The most commonly used experimental information for the determination of the parameters is the measured mean s-wave neutron resonance spacing. 
In the related experiment the A(Z,N-1) target nucleus (ground state spin $I_t$) is bombarded with low energy neutrons and the resonance spacing ($D_{l=0}$) of the A(Z,N) nucleus is determined.
The parameters $\textbf{a},\Delta,\eta $ could then be determined using the following equation:
\begin{widetext}
\[\frac{2}{{{D}_{l=0}}}=\left\{ \begin{array}{cc}
 \rho_J(S_n+\frac{1}{2}\Delta E,I_t+\frac{1}{2},\textbf{a},\Delta,\eta )+\rho_J(S_n+\frac{1}{2}\Delta E,I_t-\frac{1}{2},\textbf{a},\Delta,\eta ) \\
 \rho_J(S_n+\frac{1}{2}\Delta E,\frac{1}{2},\textbf{a},\Delta,\eta ) \\
\end{array} \right\}\begin{array}{cc}
 {{I}_{t}}\ne 0 \\
 {{I}_{t}}=0 \\
\end{array}\]
\end{widetext}
While three parameters exist, two parameters have to be determined in another way. In the case of the back-shifted Fermi gas model \cite{dilg}, the other equation is the cumulative count of the known discrete levels to be equal to the cumulative levels calculated from the level density function. 
For determination of $\eta$ one would need a third equation, which is generally non-existent. 
Therefore the level density parameters are calculated for $\eta=1.0$ and $\eta=0.5$, but there is no strong experimental evidence that any of this value is valid.

While determination of the mean s-wave neutron resonance spacing is limited to a part of the stable nucleus as target, there are only 289 nuclei with measured $D_{l=0}$ value even in the latest analysis \cite{koning2008}. 
The parameters for other nuclei are estimated based on a systematics of the known data. The data can be refined by fitting the model calculation to the experimental data.

Even early experiments and analysis had shown that the above mentioned $\eta$ values are unrealistic in some cases.
\citet{vanden} 
studied the production 
of $^{197}$Hg$^{m,g}$ isomers in two reactions, namely, $(p,n)$ and $(d,2n)$, and the isomer 
cross-section ratios 
 in $(n,\gamma$), $(d,p)$, $(n,2n)$, $(\alpha ,\alpha n)$ and $(\alpha ,xn)$ reactions.
 Based on the isomeric 
cross-section ratio, the spin-cutoff parameter 
$(\sigma )$ %
was determined and the ratio of the 
effective moment of inertia $\Theta _{eff}$ to the rigid-body moment of 
inertia $\Theta _{rig} (\eta = \Theta _{eff}/\Theta _{rig})$ was found to be about 0.1.

Our measurement and analysis \cite{sudar2006} of the pairs $^{197}$Hg$^{m,g}$ and $^{195}$Hg$^{m,g}$ has convinced that the 
$\eta$ values are in the 0.15-0.25 range for these nuclei. The $\eta$ values were also determined for $^{139}$Nd$^{m,g}$ and $^{141}$Nd$^{m,g}$ from the isomeric cross-section ratios \cite{Hilgers2007}. 
In those cases the STAPRE \cite{stro97} nuclear reaction code was used for the model calculations. In our latest evaluation \cite{Uddin2011} on $^{194}$Ir$^{m,g}$ the TALYS \cite{talys10} nuclear reaction code was used. 

 The level density parameters were determined unambiguously only in two cases: the first is the $^{51}V$ \cite{Avrig02} where the third equation for the derivation of the level density parameters was the proton resonance spacing; the second one is the $^{194}$Ir$^{m,g}$ when the isomeric cross-section ratio was the third constraint.

In this work we present the analysis for 17 new nuclei and the re-analyzed $\eta$ values of $^{197}$Hg$^{m,g}$, $^{195}$Hg$^{m,g}$, $^{139}$Nd$^{m,g}$ and $^{141}$Nd$^{m,g}$ nuclei using the procedure described in this paper. 
The data cover the mass range of 44 to 200 and our aim is to describe the mass dependence of $\eta$, i.e. the parameter $ \Theta _{eff}/\Theta _{rig}$.



\section{Method of evaluation}

The general problem in the model calculation is that the resulting cross section depends on many parameters of the model. 
The analysis of a parameter is simple when the dependence of the measured quantity on other parameters of the model is limited. 
Therefore choosing the isomeric cross-section ratio as the measured quantity is preferable because the effect of the parameters related to the incident channel may have small or negligible effect on the isomeric cross-section ratio, compared with the isomeric cross section. 
For example: if the pre-equilibrium reaction would be the dominant reaction type, the composite nucleus formation cross section would be eliminated in case of the cross-section ratio. 
Naturally, since the calculated data entail a sum of the different reaction types, there is no full cancellation of the effect of the incident channel. 
We have observed that the determination of $\eta$ from the isomeric cross-section ratio is ten times less sensitive to some of the main reaction model parameters than the isomeric cross section itself.

The isomeric cross-section ratio has another advantage from the viewpoint of reliability of the experimental data. The equations for the formation of a meta and the ground state of a product nucleus in a nuclear reaction are given below.

\begin{eqnarray} 
\label{eq1}
A_m(t)=\lambda_mN_m(t)=\lambda_m\sigma _m
n_0\Phi_{\lambda_m}e^{-\lambda_mt}
\end{eqnarray}
\begin{widetext} 
\begin{eqnarray} 
\label{eq2}
A_g(t)=\lambda_gN_g(t)=\lambda_g\left(\left(\sigma _g+
\frac{P_m\lambda_m}{\lambda_m-\lambda_g}
\sigma_m\right)
n_0\Phi_{\lambda_g}e^{-\lambda_gt}-
\frac{P_m\lambda_m}{\lambda_m-\lambda_g}
N_m(T)e^{-\lambda_mt}\right)
\end{eqnarray}
\end{widetext} 
where 
\begin{eqnarray} 
\Phi_{\lambda_x}=e^{-\lambda_xT}\int^{T}_{0}\phi(t)e^{\lambda_x t}dt
\end{eqnarray}
In this expression, $\sigma_x, \lambda_x$ are production cross section and decay constant
 for meta $(x=m)$ and ground state$(x=g)$, respectively. $P_m$ is the branching ratio of 
the metastable state to the ground state, $n_0$ the number of the target 
nuclei per $cm^2$ and $\phi(t)$ is the flux of the incident beam in particles per second.
The cross section for the metastable state can be determined directly from 
Eq. \ref{eq1}. In those cases where the half-lives of the metastable and ground states
are quite near, an analysis of the decay curve of the ground state from 
Eq. \ref{eq2} gives
 only the cumulative cross section of the metastable and the ground state, i.e.
$\sigma _c=
\sigma _g+
\frac{P_m\lambda_m}{\lambda_m-\lambda_g}
\sigma_m
$. The $\sigma _g$ could then be calculated using the data of $\sigma _c$ and $\sigma _m$. 

It is clear from the equations that the incident particle current, or neutron flux is eliminated while determining the ratio; thus an important factor of the uncertainty disappears. If one uses only Eq. \ref{eq2} for the analysis, then the uncertainty of the efficiency of the detector is also eliminated. This requires that the two components of the decay curve are well separable.

The experimental data used in the present evaluation were collected from EXFOR database \cite{EXFOR}, utilizing only those experimental data where simultaneous determination of the metastable and ground state cross sections was done or where isomeric cross-section ratio was directly reported. When only the metastable and ground state cross sections were available the uncertainty of the isomeric cross-section ratio was calculated by the formula
\begin{eqnarray} \label{eq:erreq}
\Delta \left( \frac{{{\sigma }_{m}}}{{{\sigma }_{g}}} \right)=\frac{{{\sigma }_{m}}}{{{\sigma }_{g}}}\sqrt{{{\left( \frac{\Delta {{\sigma }_{m}}}{{{\sigma }_{m}}} \right)}^{2}}+{{\left( \frac{\Delta {{\sigma }_{g}}}{{{\sigma }_{g}}} \right)}^{2}}}.
\end{eqnarray}
 This formula could overestimate the uncertainty because the uncertainty originating from the incident particle current and some other common sources are not removed.

\subsection{Model calculation}

The reaction cross sections were calculated using the nuclear model code TALYS (version 1.6), which has been recently
 developed by \citet{talys10}. It incorporates a number of nuclear models to analyze all the significant
 nuclear reaction mechanisms over the energy range of 1 keV to 200 MeV. In the calculations, the particle transmission 
coefficients were generated via the spherical optical model using the ECIS-06 code \cite{ECIS} with global parameters: for neutrons 
and protons from Koning and Delaroche \cite{koning2008}; for the optical model parameters (OMP) of complex particles (d, t, $\alpha $, $^3He$) the 
code made use of a folding approach, building up the OMPs from the neutron and proton potential. {
The TALYS always calculates the direct reaction contribution and in the case of quadrupole deformation of the target it uses the coupled channel calculation.} The gamma-ray transmission coefficients 
were calculated through the energy-dependent gamma-ray strength function according to Kopecky and Uhl \cite{JKopeczky} for E1 radiation, 
and according to Brink \cite{Brink} and Axel \cite{Axel} for all the other transition types. {
 The effect of the parameters of gamma-ray strength function on the $\eta$ values was tested and it was found that their contributions are negligible.}

For the pre-equilibrium reactions a 
two-component exciton model of the TALYS code was used. The energies, spins, parities and branching ratios of the discrete 
levels were based on the RIPL-3 database \cite{ripl3}. Since the calculated isomeric cross section is strongly dependent on the branching ratio, instead of the simple approach used in RIPL-3 to estimate the unknown branching ratio, we substituted the branching ratios, for this type of levels, by calculated data based on the gamma transmission coefficients. 
In the continuum region the level density was calculated by the back-shifted Fermi gas model (BSFG)\cite{dilg}. The TALYS uses a modified version of the BSFG model as it is in \citet{koning2008} 
\begin{eqnarray} \label{eq:a-atilde}
a=a({{E}_{x}})=\tilde{a}\left( 1+\delta W\frac{1-\exp \left[ -\gamma (E_x-\Delta) \right]}{(E_x-\Delta)} \right)
\end{eqnarray}
where $\tilde{a}$ is the asymptotic level density value which one would obtain in the absence of any shell effects, i.e.
$\tilde{a}=a({{E}_{x}}\to \infty )$ or $\delta W=0$. 
The damping parameter $\gamma $ determines how rapidly $a({{E}_{x}})$ approaches $\tilde{a}$. Finally, $\delta W$ is the shell correction energy. The excitation energy dependence of the spin-cutoff parameter is modified in two ways in the TALYS: 

1) The Eq. \ref{eq:spincut} is multiplied with the same energy dependence as applied for the level density parameter (Eq. \ref{eq:a-atilde}). This modification can be switched on/off in the input file of TALYS. We used the original energy dependence. 

2) The spin-cutoff parameter is calculated from the discrete levels and a linear interpolation applied from the mean energy of the used discrete levels to neutron separation energy, weighting discrete and BSFG spin-cutoff parameter to reach the BSFG data at the neutron separation energy. 

The spin-cutoff factor is calculated from the discrete levels in the range of N$_L$ to N$_U$ by the following equation 
\begin{eqnarray} \label{eq:spincutd}
\sigma _{d}^{2}=\frac{1}{3\sum\nolimits_{i={{N}_{L}}}^{{{N}_{U}}}{(2{{J}_{i}}+1)}}\sum\limits_{i={{N}_{L}}}^{{{N}_{U}}}{{{J}_{i}}({{J}_{i}}+1)(2{{J}_{i}}+1)},
\end{eqnarray}
where J$_i$ is the spin of the level i.
It can be shown that the TALYS has introduced an indirectly implemented energy dependence for $\eta (E_x)$ by the linear interpolation . 
The lowest value of the spin-cutoff factor seems logical and the highest value can be modified with parameters of the code. The rightness of the linear interpolation has not been proven by experiment or theory. 

Finally we excluded the continuum stripping, pick-up, break-up and knock-out reactions during the calculations because in our opinion their parametrization is not consolidated as yet. This may have some effect on $(\alpha,n)$, but not on the $(\alpha,xn)$ and $(^3He,xn)$ processes. It is also unknown how these processes feed the different spin states. 

The TALYS basically considers a limited number of the discrete levels; therefore we compared the cumulative number of the levels from the level scheme and that calculated from the level density and the highest level number $N_{dmax} $; the latter was selected until the two cumulative numbers showed good agreement.
The number of the discrete levels used is critical in the calculation of the isomeric cross-section ratios \cite{qaim88,Sudar2000} because the levels individually feed the isomeric and the ground state. 

{
The number of the discrete levels ($N_d$) is not defined in any reaction model, therefore it is a free parameter and should be determined by fitting to experimental data. 
The usual condition to determine the $N_d$ is to exclude the excitation energy range where missing levels may exist. This defines a $N_{dmax} $ value which is only an upper limit for the $N_d$  value, i. e. every $N_d \leq N_{dmax}$ fulfills the required condition. Usually, we use the $N_d = N_{dmax}$ but this condition does not give anything of the other properties of the levels as spin, parity and branching ratios. The connection between the continuum, having a wide range of spin, and the individual levels is also important. These factors are very important for the isomeric cross-section ratios. In the case of the isomeric cross-section ratios, therefore, we calculated the reduced $\chi ^2$ for every N between $N_{isomer}$ and $N_{dmax} $ and selected that N for $N_d$ which gives the lowest reduced $\chi ^2$ values. This procedure completely fulfills the requirement of the determination of the free parameter. The conditions are considered: No missing levels and best fit to the experimental data of the isomeric cross-section ratios.}

\subsection{Fitting procedure}
The general approach to find the best fit to the experimental data and the model calculation is to find the minimum of the 

\[{{\chi }^{2}}\left( \eta ,N_d \right)={{\left( \vec{e}-\vec{c}\left( \eta ,N_d \right) \right)}^{T}}{{M}^{-1}}\left( \vec{e}-\vec{c}\left( \eta ,N_d \right) \right)\]
where $\vec{e}=\left[ ...,\frac{{{\sigma }_{m}}{{\left( {{E}_{i}} \right)}_{\exp }}}{{{\sigma }_{g}}{{\left( {{E}_{i}} \right)}_{\exp }}},.. \right]$, $\vec{c}=\left[ ...,\frac{{{\sigma }_{m}}{{\left( {{E}_{i}} \right)}_{theor}}}{{{\sigma }_{g}}{{\left( {{E}_{i}} \right)}_{theor}}},... \right]$are the vectors composed of the measured and calculated isomeric cross-section ratios; ${{E}_{i}}$ is the incident energy, ${{M}^{-1}}$ the inverse of the covariance matrix of the isomeric cross-section ratio, the indices referring to experimental data or to theoretical calculation, and T refers to the transpose of the vector.

Unfortunately the covariance matrix of the isomeric cross-section ratio is usually not published; therefore the above formula can be simplified with a diagonal covariance matrix containing the uncertainties of the experimental data.
\begin{eqnarray} \label{eq:etafit}
	{{\chi }^{2}}\left( \eta ,N_d \right)=\sum\limits_{i=1}^{N}{\frac{{{\left( \frac{{{\sigma }_{m}}{{\left( {{E}_{i}} \right)}_{\exp }}}{{{\sigma }_{g}}{{\left( {{E}_{i}} \right)}_{\exp }}}-\frac{{{\sigma }_{m}}{{\left( {{E}_{i}},\eta ,N_d \right)}_{theo}}}{{{\sigma }_{g}}{{\left( {{E}_{i},\eta ,N_d } \right)}_{theo}}} \right)}^{2}}}{{{\varepsilon }^{2}}{{\left( {{E}_{i}} \right)}_{\exp }}}}
\end{eqnarray}
where ${{E}_{i}}$ is the incident energy, $\varepsilon $ the uncertainty of the isomeric cross-section ratio, and the indices refer to experimental data or theoretical calculation.

While we have at least two free parameters, the calculation for a series of both parameters has to be repeated several times, until the minimum value converges to a constant. On the related figures below are shown these final results. Usually it would need two or three runs per parameter.

The TALYS has two methods to set the $\eta$ parameter of the model: 

1) Using the \textbf{Rspincut} keyword of the code, which defines the same $\eta$ value for all nuclei treated in the calculation. 

2) Using the \textbf{s2adjust} keyword of the code, which defines the $\eta$ values for only one nucleus specified in the parameters of the keyword.

When the evaluation was started it was supposed that the mass dependence of the $\eta$ is weak, therefore the use of the \textbf{Rspincut} keyword was appropriate. The result of the evaluation has shown an even-odd effect in the data. This required  testing and finally analyzing the effect of the neighboring nuclei in multistage reactions.

The ${\chi }^{2}\left( \eta ,N_d \right)$ plot gives a possibility to estimate the uncertainty of the $\eta $ value \cite{Wall1996}.

\subsection{Analysis of the $\eta $ value from the discrete levels}

In the case of low mass numbers, the excitation schemes of several nuclei are known even up to high energy.
This gives a possibility to test the energy dependence of the $\eta $. We calculated $\eta $ for the discrete levels, using a method similar to that by \citet {Al-Quraishi} who fitted the spin-cutoff factor by the following formula for the discrete levels:
\begin{eqnarray} \label{Alq} 
\sigma=constA^{5/6}(U-\Delta)^{1/4}/\textbf{a}^{1/4},
\end{eqnarray}
 for 
the energy(U), energy shift($\Delta$), mass number(A) and level density parameter(\textbf{a}) 
dependence of $\sigma$. The fitted "const" in the formula showed mass number dependence.

Comparing the equation above with Eq. \ref{eq:spincut}, their "const" can be connected to $\eta $ but the simplest is to substitute $\sigma_d$ in Eq. \ref{eq:spincut} and expressing $\eta $ from the equation we get:
\begin{eqnarray} \label{eq:deta}
\eta_d=\frac{{\sigma_d}^2}{\frac{\Theta _{rig}}{\hbar ^2}\sqrt{\frac{E_x-\Delta}{\textbf{a}}}}
\end{eqnarray}
The energy range of the discrete levels is divided into energy grid of 0.5 MeV, the ${\sigma_d} ^2$ is calculated by the Eq. \ref{eq:spincutd} and $\eta_d $ is calculated for each interval by Eq.~\ref{eq:deta}. 
To remove the fluctuation, a moving average for $\eta $ using three consecutive intervals was also calculated. The energy shift($\Delta$) and level density parameter(\textbf{a}) were taken from the evaluation of \citet {Al-Quraishi}. Their data refer to the low excitation energy, and the parameters were calculated from the discrete levels.

\section{Results}

\subsection{$^{46}$Ti$(\alpha ,x)^{44}$Sc$^{m,g}$}

\begin{figure}
\centering\includegraphics*[scale=\ones ]{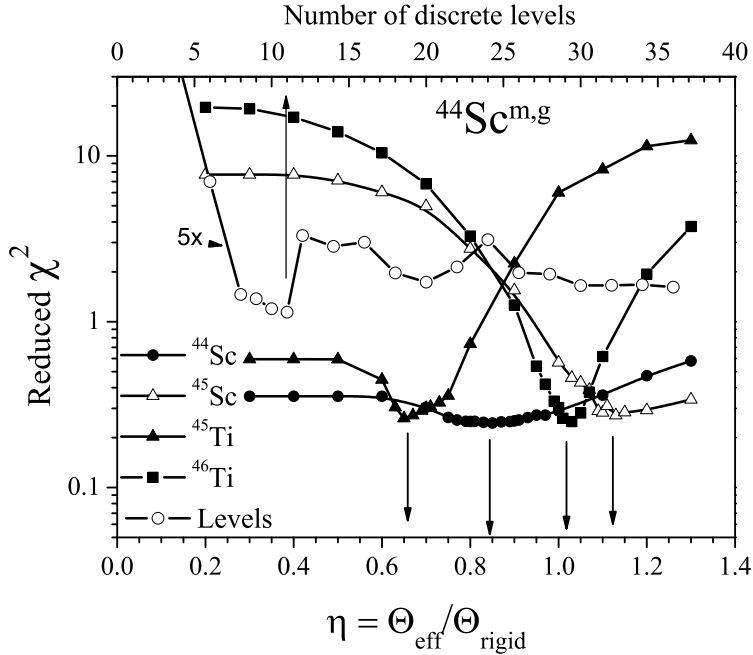}
\caption{\label{fig:44Sc} Reduced $\chi ^2$ as a function of the number of discrete levels (multiplied by 5) and $\eta $ for the $^{44}$Sc, $^{45}$Sc, $^{45}$Ti and $^{46}$Ti nuclei based on the isomeric cross-section ratios of the $^{46}$Ti$(\alpha ,x)^{44}$Sc$^{m,g}$ reaction.
}
\end{figure}

\begin{figure}
\centering\includegraphics*[scale=\ones ]{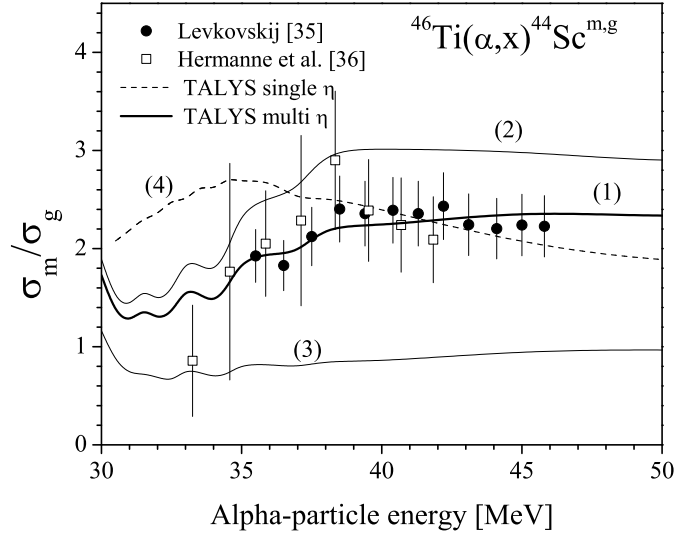}
\caption{\label{fig:44Scmg} Measured and calculated isomeric cross-section ratios for the $^{46}$Ti$(\alpha ,x)^{44}$Sc$^{m,g}$ reaction, using the optimal number of discrete levels and the best $\eta$ values on curve marked with (1), while curves (2) and (3) are based on $\eta$ values of 1.0 and 0.50, respectively, of $^{46}$Ti. The curve (4) shows the best value of the calculation using the same $\eta$ value for all nuclei.  
}
\end{figure}

\begin{figure}
\centering\includegraphics*[scale=\ones ]{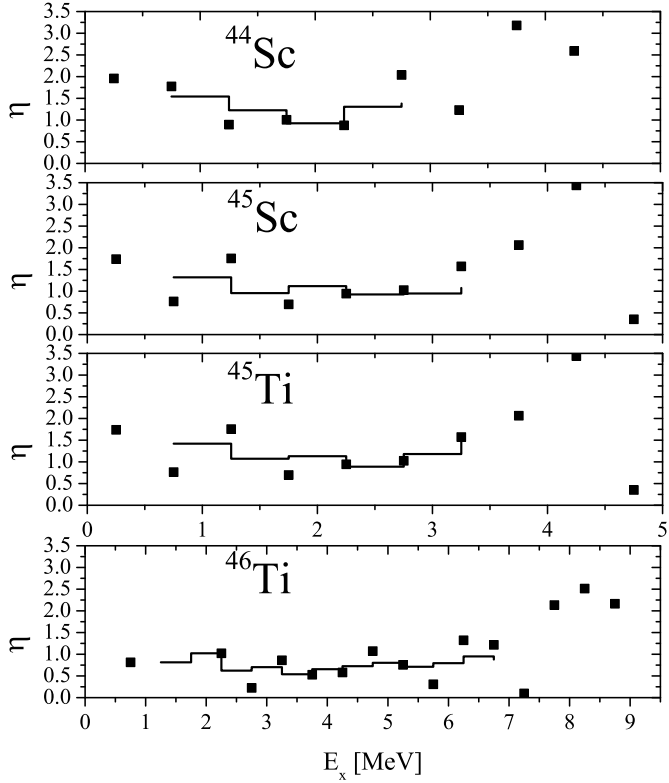}
\caption{\label{fig:44Scd1} $\eta $ values calculated from the discrete levels of the $^{44}$Sc, $^{45}$Sc, $^{45}$Ti and $^{46}$Ti nuclei. The points show the $\eta $ values for 0.5 MeV bins, while the continuous line depicts the moving average of three bins, indicating the tendency of the energy dependence better.
}
\end{figure}

\begin{figure}
\centering\includegraphics*[scale=\ones ]{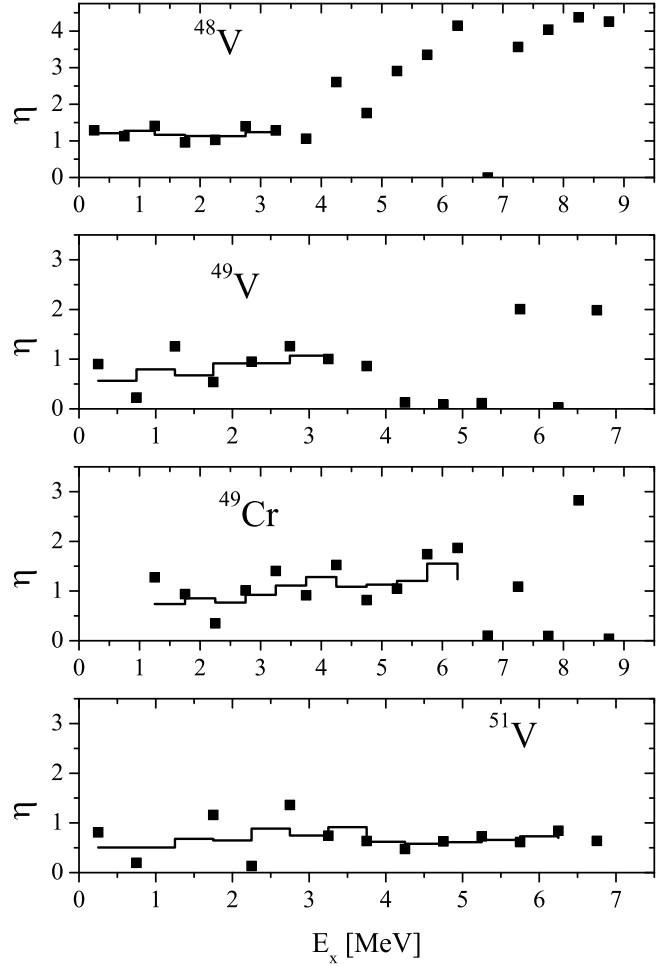}
\caption{\label{fig:44Scd2}
$\eta $ values calculated from the discrete levels of the $^{48}$V, $^{49}$V, $^{49}$Cr and $^{51}$V nuclei. 
The marking has the same meaning as in Fig.~\ref{fig:44Scd1}. 
}
\end{figure}
\begin{table*}
\caption[]{\label{tab:Sc-V-cd}Summary of the evaluated continuum and discrete $\eta $ from $^{46}$Ti$(\alpha ,apn)^{44}$Sc$^{m,g}$, $^{51}$V$(\alpha ,3n)^{52}$Mn$^{m,g}$, $^{nat}$Cr$(^3He,3n)^{52,53}$Fe$^{m,g}$ and $^{55}$Mn$(\alpha ,n)^{58}$Co$^{m,g}$ reactions and ENSDF \cite{ensdf} database.
}
\begin{tabular}{ccccccccccccccc}
\hline
&&&Cont.&&&&Disc.\\
Nucleus & A & Z & $\eta $ & $\Delta \eta $ & E$_x$ low & E$_x$ max & $\eta _d$ & $\Delta \eta _d$ & E$_x$ & $\eta $/$\eta _d$ & $\Delta$($\eta $/$\eta _d$) & S$_n$ & S$_p$ \\
\hline
$^{44}$Sc & 44 & 21 & 0.85 & 0.27 & 0.00 & 24.66 & 0.93 & 0.07 & 1.5-3 & 0.92 & 0.30 & 9.699 & 6.696 & ($\alpha $,x)\\
$^{45}$Sc & 45 & 21 & 1.13 & 0.17 & 11.33 & 35.99 & 1.07 & 0.42 & 0-3.5 & 1.06 & 0.45 & 11.33 & 6.891 & ($\alpha $,x)\\
$^{45}$Ti & 45 & 22 & 0.65 & 0.15 & 8.48 & 33.14 & 1.15 & 0.47 & 0-3 & 0.56 & 0.27 & 9.531 & 8.482 & ($\alpha $,x)\\
$^{46}$Ti & 46 & 22 & 1.03 & 0.07 & 21.66 & 41.62 & 0.79 & 0.36 & 0-7 & 1.30 & 0.59 & 13.18 & 10.34 & ($\alpha $,x)\\
$^{48}$V & 48 & 23 & 0.91 & 0.4 & 9.08 & 33.74 & 1.21 & 0.18 & 0-3.5 & 0.75 & 0.35 & 10.54 & 6.829 & ($\alpha $,x)\\
$^{49}$V & 49 & 23 & 1.00 & 0.12 & 20.64 & 45.30 & 0.88 & 0.38 & 0-3.5 & 1.14 & 0.51 & 11.55 & 6.758 & ($\alpha $,x)\\
$^{49}$Cr & 49 & 24 & 0.99 & 0.13 & 17.23 & 41.89 & 1.21 & 0.40 & 4-6 & 0.82 & 0.29 & 10.58 & 8.145 & ($\alpha $,x)\\
\hline 
$^{51}$V & 51 & 23 & 0.75\footnotemark[1] & 0.06\footnotemark[1] & 10.344 & 13.18 & 0.66 & 0.12 & 4-7.5 & 1.14 & 0.23 & 13.18 & 10.34 & \\
\hline 
$^{52}$Mn & 52 & 25 & 1.05 & 0.07 & 5.58 & 20.28 & 1.14 & 0.35 & 0-5.5 & 0.92 & 0.29 & 10.53 & 6.54 & ($\alpha $,3n)\\
$^{53}$Mn & 53 & 25 & 1.00 & 0.10 & 17.63 & 32.33 & 0.66 & 0.31 & 0-4.5 & 1.52 & 0.73 & 12.05 & 6.56 & ($\alpha $,3n)\\
$^{54}$Mn & 54 & 25 & 1.17 & 0.11 & 26.57 & 39.89 & 1.03 & 0.47 & 0-3.5 & 1.13 & 0.53 & 8.94 & 7.56 & ($\alpha $,3n)\\
\hline 
$^{52}Fe$ & 52 & 26 & & & 0.00 & 16.88 & 0.71 & 0.25 & 2.5-7& & & 16.20 & 7.37 \\
$^{53}Fe$ & 53 & 26 & 0.97\footnotemark[2] & 0.06\footnotemark[2] & 10.69 & 27.57 & 0.84 & 0.41 & 0.5-6 & 1.16 & 0.57 & 10.69 & 7.53 & ($^3He $,xn) \\
$^{54}Fe$ & 54 & 26 & 0.58\footnotemark[1] & 0.06\footnotemark[1] & 24.06 & 42.67 & 0.51	&	0.17	&	3-6.5	&	1.14	&	0.4 & 13.38 & 8.85 & ($^3He $,xn)\\
$^{54}Fe$ & 54 & 26 & 0.53\footnotemark[3] & 0.02\footnotemark[3] & 13.37 & 42.67 & 0.51	&	0.17	&	3-6.5	 & 1.04	&	0.35 & 13.38 & 8.85 & ($^3He $,xn)\\
$^{55}Fe$ & 55 & 26 & 0.81\footnotemark[3] & 0.16\footnotemark[3]& 9.298 & 33.11 & 0.46 & 0.33 & 0-3.5 & 1.77 & 1.32 & 9.30 & 9.21 & ($^3He $,xn)\\
$^{58}Co$ & 58 & 27 & 1.02\footnotemark[4] & 0.13\footnotemark[4] & 0.00 & 22.71 & 0.66 & 0.13 & 0-1.5 & 1.54 & 0.37 & 8.57 & 6.95 & ($\alpha $,n) \\
\hline

\end{tabular}
\begin{flushleft}
\footnotemark[1] reported $\eta $ value of \citet{Avrig02}

\footnotemark[2] $\eta $ value from $^{nat}$Cr$(^3He ,xn)^{52}$Fe$^{m,g}$ reaction

\footnotemark[3] $\eta $ value from $^{nat}$Cr$(^3He ,xn)^{53}$Fe$^{m,g}$ reaction

\footnotemark[4] $\eta $ value from $^{55}$Mn$(\alpha ,n)^{58}$Co$^{m,g}$ reaction
\end{flushleft}
\end{table*}
The experimental isomeric cross-section ratios were calculated from the cross section data of \citet{Levk91} and \citet{Herm99}. The formula \ref{eq:erreq} was used to calculate their uncertainties.
Before the model calculation, the level scheme of the product $^{44}$Sc  in the TALYS input files was checked, and the simple approximation for the unknown branching ratios was substituted by the calculated one using the gamma transmission coefficient. Both the optimal number of the used discrete levels and $\eta$ were determined with the procedure described above. Fig.~\ref{fig:44Sc} shows the reduced $\chi ^2$ of the fitting of the model-calculated data to the experimental data as a function of the used discrete levels and $\eta$ values. 
The minimum value of the ${\chi }^{2}$ corresponds to the optimal number of levels amounting to 11. Thus, in this case, only one level with unknown branching ratio was used (725 keV). 
The isomeric cross-section ratio as a function of the $(\eta = \Theta _{eff}/\Theta _{rig})$ was calculated and compared with the experimental data. Only the $^{44}$Sc, $^{45}$Sc, $^{45}$Ti, $^{46}$Ti, $^{48}$V, $^{49}$V and $^{49}$Cr were treated as intermediate nuclei in the iteration process, i.e. one neutron, a proton and an alpha particle emission were handled. 
The intermediate nuclei originating from the substitution of the alpha particle by 2n2p was neglected in the analysis of the $\eta$ because of the higher reaction threshold. 
Fig.~\ref{fig:44Sc} shows only the $\eta$ dependence for the $^{44}$Sc, $^{45}$Sc, $^{45}$Ti and $^{46}$Ti nuclei while the reduced ${\chi }^{2}$ of the $^{48}$V, $^{49}$V and $^{49}$Cr overlapped each other. 
Table \ref{tab:Sc-V-cd} presents a summary of the evaluated data with the estimated excitation energy range effective in the evaluation. Fig.~\ref{fig:44Scmg} shows the measured and calculated isomeric cross-section ratios as a function of the incident $\alpha $-particle energy for four different cases: 
(1) The best fit to the experimental data can be achieved by $\eta $ values presented in Table ~\ref{fig:44Scmg}. Curves (2) and (3) are based on $\eta$ values of 1.0 and 0.50, respectively, of $^{46}$Ti, whose $\eta$ value is the most sensitive for calculating the isomeric cross-section ratio. The curve (4) depicts the best value of the calculation using the same $\eta$ value for all nuclei. 
Comparing the curves (1) and (4), it is clear that local differences in the $\eta $ values play an important role in the description of the isomeric cross-section ratio. 

The discrete levels of the nuclei mentioned above are known up to relatively high excitation energy; therefore, we evaluated the $\eta_d$ values from the discrete levels using the Eq. \ref{eq:deta} and Eq. \ref{eq:spincutd} with 0.5 MeV bin size. 
The results are shown in Fig.~\ref{fig:44Scd1} and Fig.~\ref{fig:44Scd2}. They provide the possibility to get information on the energy dependence of the $\eta_d $ values, at least in the low energy region. 
The points represent the calculated $\eta_d $ values for the 0.5 MeV bins and the continuous line is the moving average for the three bins. It is a known problem with the discrete levels that at higher energies levels are missing from the level scheme. 
While higher levels can be identified through the coincidence of the gamma transitions with lower gammas, the spins of those levels are increasing as usual in the gamma cascade. 
So the missing levels have  low spins. Therefore sharply rising $\eta_d $ values indicate the dominance of the lost levels. 
The moving average was calculated only for those excitation energy regions where the effect of the missing levels seems negligible. We have included into the analysis $^{51}$V which is in this mass number range and for which $\eta $ value has been reported  by \citet{Avrig02}. 
The $\eta_d $ value of $^{45}$Sc, $^{45}$Ti, $^{48}$V and $^{51}$V nuclei seems to be constant. The $\eta_d $ value of $^{44}$Sc shows a decreasing trend with the increasing energy in the range of 0-2 MeV, but there is an increasing trend of $\eta_d $ with the increasing energy for the nuclei $^{49}$V and $^{49}$Cr. 
In the case of $^{46}$Ti, the $\eta $ value decreases up to 4 MeV and then slightly increases up to 7 MeV. The estimated average $\eta_d $ values in the appropriate energy interval are given in  Table~\ref{tab:Sc-V-cd} which also includes the calculated ratio of evaluated $\eta $ values to the data obtained from the discrete levels ( $\eta $/$\eta _d$) and also its uncertainty. 
The weighted average of the $\eta $/$\eta _d$ ratio is $0.91\pm 0.12$. (The weight is the inverse square of its uncertainty.) 
This average indicates a similar behavior of the levels in the continuum as the discrete ones. 
It is worth noting that in the case of $^{51}$V both the discrete levels up to 7.5 MeV and the direct determination of $\eta $ give the same value of about 0.75, indicating that this lower value may originate from the nuclear structure.

%
\subsection{$^{51}$V$(\alpha ,3n)^{52}$Mn$^{m,g}$}

\begin{figure}
\centering\includegraphics*[scale=\ones ]{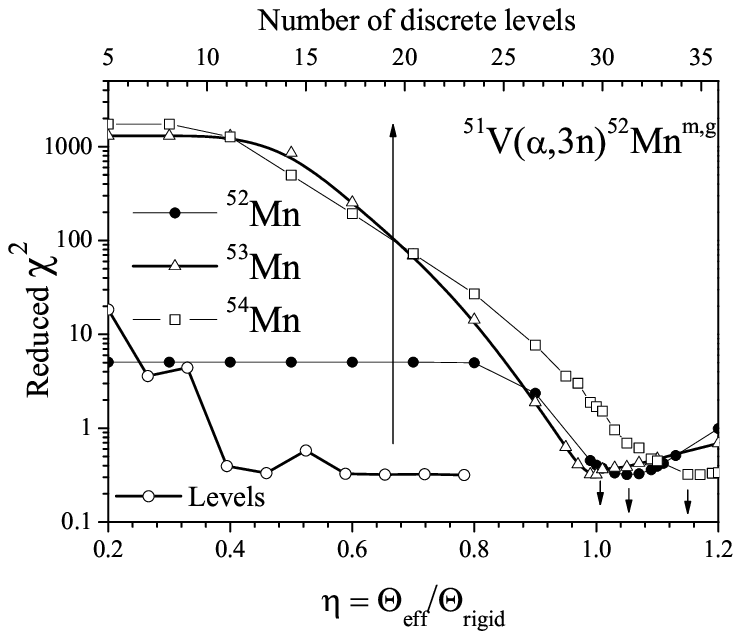}
\caption{\label{fig:52Mn} Reduced $\chi ^2$ as a function of the number of discrete levels and $\eta $ for the $^{51}$V$(\alpha ,3n)^{52}$Mn$^{m,g}$ reaction. 
}
\end{figure}
\begin{figure}
\centering\includegraphics*[scale=\ones ]{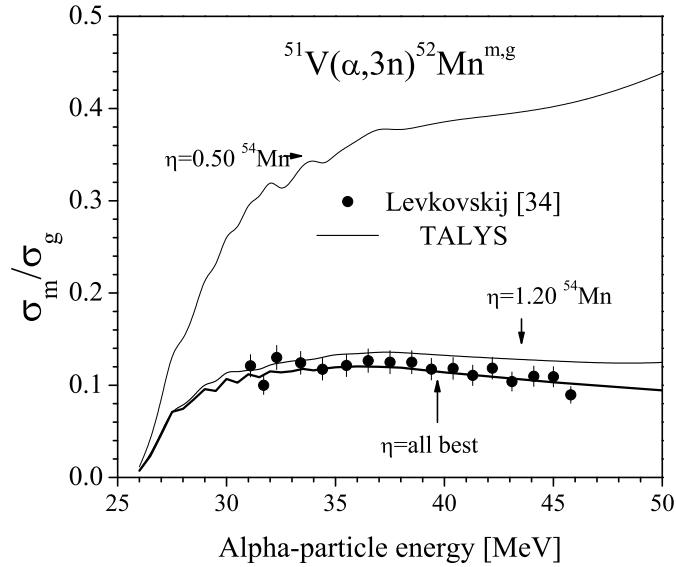}
\caption{\label{fig:52Mnmg} Measured and calculated isomeric cross-section ratios for the $^{51}$V$(\alpha ,3n)^{52}$Mn$^{m,g}$reaction using the optimal number of discrete levels and the best $\eta $ values of fit as well as $\eta$ values of 1.20 and 0.50 of $^{54}$Mn. 
}
\end{figure}

\begin{figure}
\centering\includegraphics*[scale=\ones ]{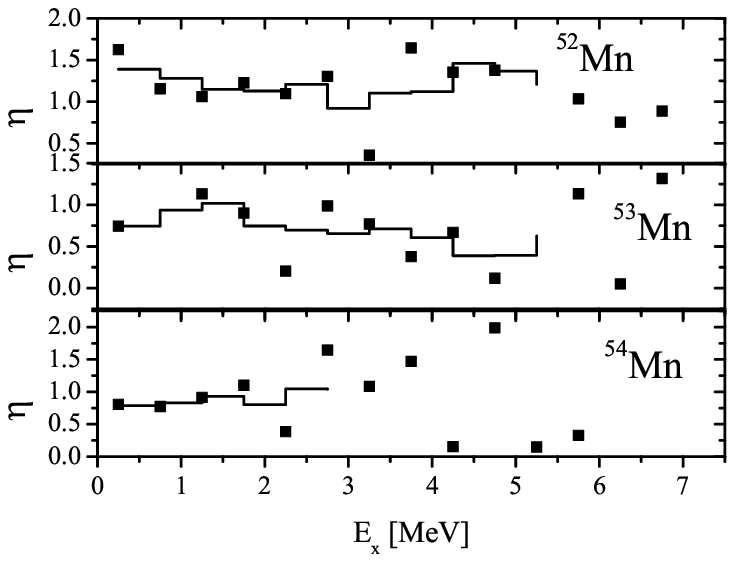}
\caption{\label{fig:52Mnd1} $\eta $ values calculated from the discrete levels of the $^{52}$Mn, $^{53}$Mn and $^{54}$Mn nuclei. The points show the $\eta $ value for 0.5 MeV bins, while the continuous line depict moving average of three bins to show the tendency of the energy dependence better.
}
\end{figure}

The experimental isomeric cross-section ratios were calculated from the cross section data of \citet{Levk91}. The formula \ref{eq:erreq} was used to calculate their uncertainties. 
This method overestimates the uncertainties because the error of the beam current should be eliminated from the final result. 
The branching ratios of the levels in $^{52}$Mn in the TALYS library were examined, and the simple approximation was substituted by our approach. The change was necessary only for the level 22 ($2.645~MeV$) and above. Fig.~\ref{fig:52Mn} shows the reduced $\chi ^2$ for comparing the model-calculated data with the experimental data. 
The best fit to the experimental data as a function of the number of the levels (N) is obtained at N=19, but in this case, there was only a small change in the reduced $\chi ^2$ above level 19. 
Fig.~\ref{fig:52Mn} also depicts the reduced $\chi ^2$ as a function of the $\eta$ parameter calculated after fixing the number of the levels. 
To get the best fit for the $^{51}$V$(\alpha ,3n)^{52}$Mn$^{m,g}$ reaction the required $\eta$ values for the $^{52}$Mn, $^{53}$Mn and $^{54}$Mn isotopes are summarized in the Table~\ref{tab:Sc-V-cd}. 
It seems rather curious that $\eta$ values of the $^{52}$Mn and $^{54}$Mn are higher than 1, although the $\eta $ is the ratio of $ \Theta _{eff}$ to $\Theta _{rig}$ which could be less or equal to 1. The value of the nuclear radius parameter has an effect on the  $\Theta _{rig}$ value. Otherwise taking into account the uncertainty of the $\eta$, those values can be accepted. 
Fig.~\ref{fig:52Mnmg} depicts the measured and calculated isomeric cross-section ratios as a function of the incident $\alpha $-particle energy with three different $(\eta = \Theta _{eff}/\Theta _{rig})$ values (0.50 for $^{54}$Mn, best fit for all isotopes, 1.20 for $^{54}$Mn ). 
Table~\ref{tab:Sc-V-cd} presents the $\eta $ values giving the best fit to the experimental data.

The discrete levels of the $^{52}$Mn, $^{53}$Mn and $^{54}$Mn nuclei are also known up to relatively high energy; therefore, we have evaluated the $\eta _d$ values from the discrete levels using the Eq. \ref{eq:deta} and Eq. \ref{eq:spincutd} as above. The results are shown in Fig.~\ref{fig:52Mnd1}. 
The symbols represent the calculated $\eta _d$ values for the 0.5 MeV bins and the continuous lines are the moving averages for the three bins. 
The moving average seems approximately constant for $^{52}$Mn and $^{54}$Mn near to value of 1.0, but for $^{53}$Mn it shows a decreasing tendency up to 4.5 MeV. 
The Table~\ref{tab:Sc-V-cd} gives estimated average $\eta _d$ values in the appropriate energy interval, and the calculated ratio of evaluated $\eta $ values, based on the isomeric cross-section ratio, to their data obtained from the discrete levels ( $\eta $/$\eta _d$) and its uncertainty too. 
It seems that in the case of $^{53}$Mn the $\eta $ value obtained at higher excitation energy is near 1.0 but it  is only 0.66 from the discrete levels. This difference does not indicate disagreement taking into account the uncertainties of the data.

%

\subsection{$^{nat}$Cr$(^3He ,xn)^{52}$Fe$^{m,g}$ and $^{nat}$Cr$(^3He ,xn)^{53}$Fe$^{m,g}$}

\begin{figure}
\centering\includegraphics*[scale=\ones ]{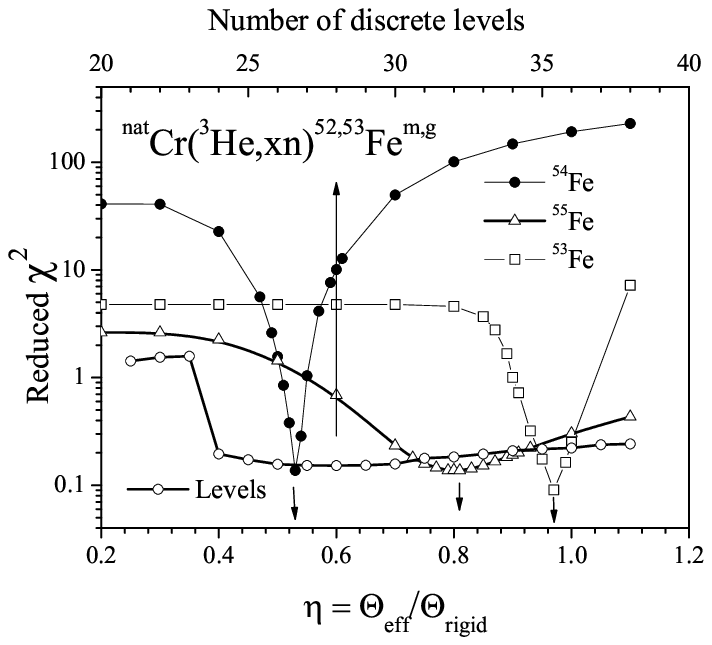}
\caption{\label{fig:53Fe}Reduced $\chi ^2$ as a function of the number of discrete levels and $\eta $ for the $^{nat}$Cr$(^3He ,xn)^{52,53}$Fe$^{m,g}$ reaction.
}
\end{figure}
\begin{figure}
\centering\includegraphics*[scale=\ones ]{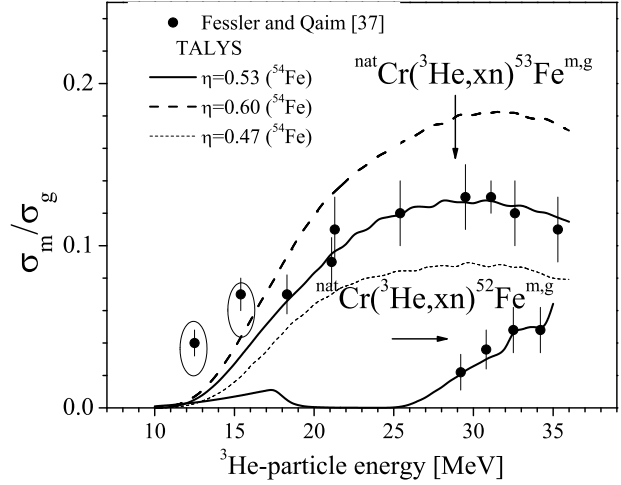}
\caption{\label{fig:53Femg} Measured and calculated isomeric cross-section ratios for the $^{nat}$Cr$(^3He ,xn)^{52,53}$Fe$^{m,g}$ reaction, using the optimal number of discrete levels and the $\eta $ values of 0.53, 0.60 and 0.47 for $^{54}$Fe.   The calculation  is shown only the result with the best $\eta $ values for the production of the $^{52}$Fe$^{m,g}$. 
}
\end{figure}

\begin{figure}
\centering\includegraphics*[scale=\ones ]{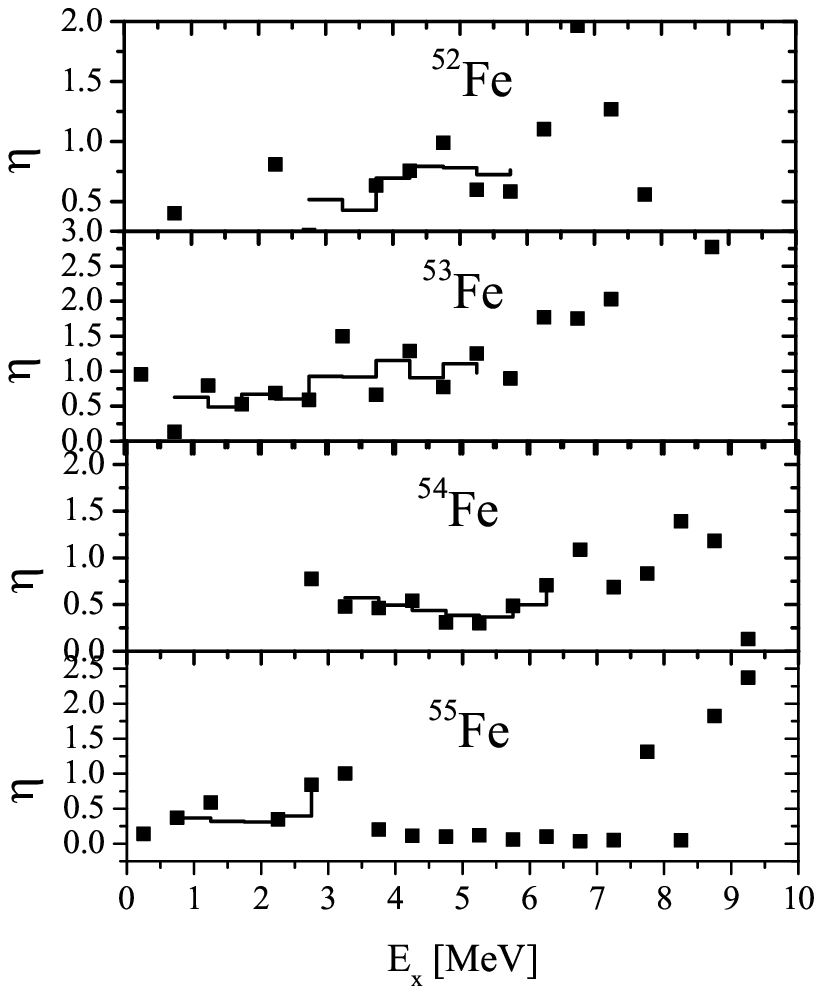}
\caption{\label{fig:52-55Fed1} $\eta $ values calculated from the discrete levels of the $^{52}$Fe, $^{53}$Fe, $^{54}$Fe and $^{55}$Fe nuclei. The points show the $\eta $ values for 0.5 MeV bins, while the continuous line depicts moving average of three bins.
}
\end{figure}

These cases have some specialty: two neighboring isomeric nuclei excited in the same reaction and the excitation energy range of the other nuclei, involved in the reaction chain, are different. 
Therefore we could get direct information on the energy dependence of $\eta $ at high excitation energy. 
The isomeric states of $^{52}$Fe and $^{53}$Fe have high excitation energy ($^{52}$Fe 6.82 MeV and $^{53}$Fe 3.04 MeV), therefore continuum has a small role in the population of the isomeric and ground state. 
The $\eta $ values of $^{52}$Fe and $^{53}$Fe have negligible effect on its isomeric cross-section ratio, but $\eta $ values of $^{53}$Fe can be determined from isomeric cross-section ratio of $^{52}$Fe.

The experimental isomeric cross-section ratios were calculated from the cross section data of \citet{fessler1996}. The branching ratios of the levels in $^{52}$Fe and $^{53}$Fe in the TALYS library were examined, and the simple approximation was substituted by our approach. 
Fig~\ref{fig:53Fe} depicts the reduced $\chi ^2$ for comparing the model-calculated data with the experimental data. The curve for $^{54}$Fe shows the calculated data for the $^{nat}$Cr$(^3He ,xn)^{53}$Fe$^{m,g}$ reaction only. The data  for the $^{nat}$Cr$(^3He ,xn)^{52}$Fe$^{m,g}$ reaction  are quite near to the other one, and it would make the figure unclear.
 The best fit to the experimental data as a function of the number of the levels (N) is obtained at N=28 for $^{53}$Fe, but in this case there was only a small change in the reduced $\chi ^2$ above level 28. 
 Fig.~\ref{fig:53Fe} also depicts the reduced $\chi ^2$ as a function of the $\eta$ parameter calculated after fixing the number of the levels. To get the best fit on $^{nat}$Cr$(^3He ,xn)^{52,53}$Fe$^{m,g}$ reaction the required $\eta$ values for the $^{53}$Fe, $^{54}$Fe and $^{55}$Fe isotopes are summarized in the Table~\ref{tab:Sc-V-cd}. 
 Fig.~\ref{fig:53Femg} depicts the measured and calculated isomeric cross-section ratios as a function of the incident $^3He $-particle energy for both $^{nat}$Cr$(^3He ,xn)^{52}$Fe$^{m,g}$ and $^{nat}$Cr$(^3He ,xn)^{53}$Fe$^{m,g}$ using the fitted $\eta $ values. 
 The two low incident energy data of $^{nat}$Cr$(^3He ,xn)^{53}$Fe$^{m,g}$ reaction seem to be shifted in energy, therefore those two values were not considered in the fitting procedure.
 
The discrete levels of the $^{52}$Fe, $^{53}$Fe, $^{54}$Fe and $^{55}$Fe nuclei are also known up to relatively high excitation energy; therefore, we have evaluated the $\eta _d$ values from the discrete levels using the Eq. \ref{eq:deta} and Eq. \ref{eq:spincutd} as above. 
The results are shown in Fig.~\ref{fig:52-55Fed1}. The points represent the calculated $\eta _d$ values for the 0.5 MeV bins and the continuous line is the moving average for the three bins. The $\eta _d$ values from the discrete levels seem to be constant in the case of the $^{52}$Fe, $^{54}$Fe and $^{55}$Fe nuclei but show a slightly increasing tendency in the case of $^{53}$Fe. 
The ratio of $\eta $/$\eta _d$ is near to one in the estimated uncertainty limits. The most significant result is in the case of $^{54}$Fe where there are three data sets in three different energy ranges. 
We can state that the $\eta $ value is constant up to 20 MeV excitation energy or even above, with the value near to 0.5. Thus the $\eta $ value seems to be constant at high excitation energy.

\subsection{$^{55}$Mn$(\alpha ,n)^{58}$Co$^{m,g}$}

\begin{figure}
\centering\includegraphics*[scale=\ones ]{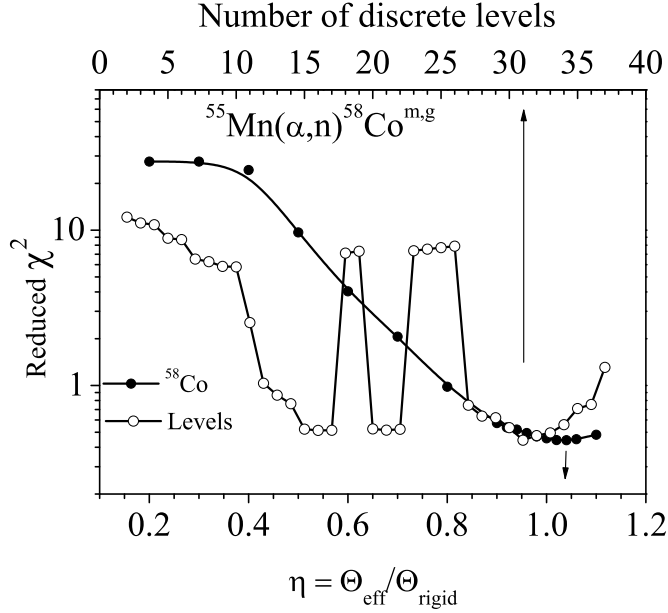}
\caption{\label{fig:58Co} Reduced $\chi ^2$ as a function of the number of discrete levels and $\eta $ for the $^{55}$Mn$(\alpha ,n)^{58}$Co$^{m,g}$ reaction. 
}
\end{figure}

\begin{figure}
\centering\includegraphics*[scale=\ones ]{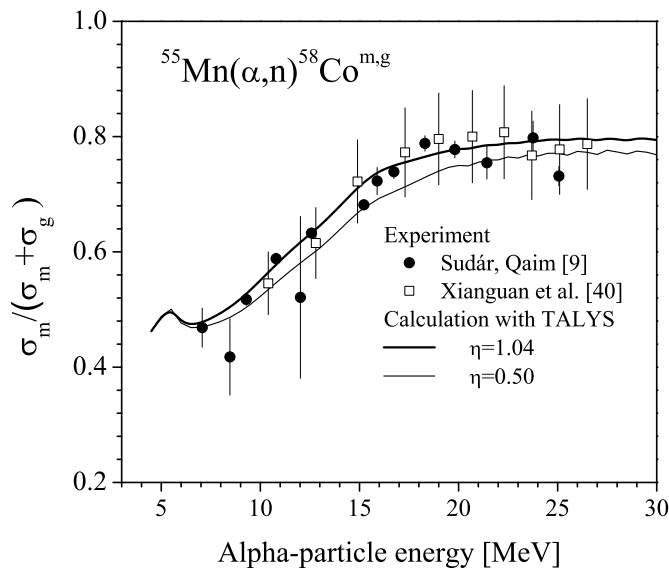}
\caption{\label{fig:58Comg} Measured and calculated isomeric cross-section ratios for the $^{55}$Mn$(\alpha ,n)^{58}$Co$^{m,g}$reaction using the optimal number of discrete levels and the $\eta $ values of 1.04 and 0.50.
}
\end{figure}

The isomeric cross-section ratio for production of the $^{58}$Co$^{m,g}$ pair was investigated earlier in many works \cite{sudar96,Avrig99,semkova04}. In the case of the $^{55}$Mn$(\alpha ,n)^{58}$Co$^{m,g}$ reaction the 
\mbox{ $\sigma_m/(\sigma_g+\sigma_m)$} was compared with the model calculation because these data can be derived directly from the experiment. It was shown by \citet{sudar96} that there is a discrepancy in the decay mode of the third level of $^{58}$Co but in spite of the four different reactions investigated, the discrepancy could not be resolved. 
\citet{Avrig99} determined the transition rate of the third level to the ground and the first isomeric state using the measured isomeric cross-section ratios from the (n,p) reaction, from the cross section measured by relatively low energy neutrons, when only a few known discrete levels of $^{58}$Co were excited. 
We used those data in the present evaluation. The experimental isomeric cross-section ratios were selected from \citet{sudar96} and \citet{Xianguan90}. 
Fig.~\ref{fig:58Co} depicts the reduced $\chi ^2$ for comparing the model-calculated data with the experimental data. The best fit to the experimental data as a function of the number of the levels (N) is obtained at N=31. This figure shows the importance of the selection of the appropriate number of levels: 
The reduced $\chi ^2$ sharply fluctuates with the number of used levels. Thus a random selection of the number of levels can result in a bad fit to the experimental data. Here we not only show the problem, which has been mentioned in many publications, but we are presenting a method how to solve it. 
The reduced $\chi ^2$ shown as a function of the $\eta $ parameter has a definite minimum at $\eta =1.04\pm 0.08$. That the minimum is not too sharp is reflected in the larger uncertainty. 
Fig.~\ref{fig:58Comg} shows the measured and calculated isomeric cross-section ratios as a function of the incident $\alpha $-particle energy with two different $\eta $ values (0.50 and 1.04). 

%

\subsection{$^{73}$Se$^{m,g}$}

The experimental isomeric cross-section ratios of $^{73}$Se$^{m,g}$ were studied in detail for the reactions $^{70}$Ge$(\alpha ,n)^{73}$Se$^{m,g}$, $^{nat}$Ge$(^3He ,xn)^{73}$Se$^{m,g}$, $^{74}$Se$(n,2n)^{73}$Se$^{m,g}$, $^{75}$As$(p,3n)^{73}$Se$^{m,g}$ and $^{75}$As$(d,4n)^{73}$Se$^{m,g}$ by \citet{qaim88}. 
The model calculations were carried out by the nuclear reaction code STAPRE for $\eta =0.50$ and $\eta =1.0$. It was found that the level scheme of $^{73}$Se has to be modified to be able to describe the excitation functions and the isomeric cross-section ratios relatively well by the model calculation. 
Especially it was necessary to remove the third level (at 26.4 keV, $5/2$ ) and all transitions that would feed this level were set to the isomeric state at 25.7 keV. 
We reevaluate the data in the framework used in this paper to determine more precisely the $\eta $ parameter.  The $^{70}$Ge$(\alpha ,n)^{73}$Se$^{m,g}$ reaction has been removed from our studies based on the unconsolidated problem of direct reactions to the continuum.  
Each reaction is considered individually.

\begin{table*}
\centering
\caption{\label{tab:Se73-76}Summary of the evaluated continuum and discrete $\eta $ from different reactions producing $^{73}$Se and ENSDF \cite{ensdf} database.}
\begin{tabular}{cccccccccccccccc}
\hline
&&&Cont.&&&&Disc.&\\
Nucleus & A & Z & $\eta $ & $\Delta \eta $ & E$_x$ low & E$_x$ max & $\eta _d$ & $\Delta \eta _d$ & E$_x$ & $\eta $/$\eta _d$ & $\Delta$($\eta $/$\eta _d$) & S$_n$ & S$_p$ &	Reaction\\
\hline
$^{73}$Se	&	73	&	34	&	0.78	&	0.48	&	0.00	&	5.97	&	0.87	&	0.13	&	0.5-1.5	&	0.90	&	0.57	&	8.43	&	7.29	&	(n,2n)	\\
$^{73}$Se	&	73	&	34	&	0.81	&	0.12	&	0.00	&	20.25	&	0.87	&	0.13	&	0.5-1.5	&	0.93	&	0.20	&	8.43	&	7.29	&	(d,4n)	\\
$^{73}$Se	&	73	&	34	&	0.65	&	0.07	&	0.00	&	22.83	&	0.87	&	0.13	&	0.5-1.5	&	0.75	&	0.14	&	8.43	&	7.29	&	(p,3n)	\\
$^{73}$Se	&	73	&	34	&	0.93	&	0.22	&	0.00	&	36.47	&	0.87	&	0.13	&	0.5-1.5	&	1.07	&	0.30	&	8.43	&	7.29	&	(He$^3$,xn)	\\
$^{74}$Se	&	74	&	34	&	0.47	&	0.4	&	12.06	&	18.03	&	0.46	&	0.16	&	0.5-6	&	1.01	&	0.93	&	12.06	&	8.55	&	(n,2n)	\\
$^{74}$Se	&	74	&	34	&	0.43	&	0.08	&	12.06	&	32.31	&	0.46	&	0.16	&	0.5-6	&	0.93	&	0.36	&	12.06	&	8.55	&	(d,4n)	\\
$^{74}$Se	&	74	&	34	&	0.59	&	0.16	&	12.06	&	34.89	&	0.46	&	0.16	&	0.5-6	&	1.27	&	0.55	&	12.06	&	8.55	&	(p,3n)	\\
$^{74}$Se	&	74	&	34	&	0.43	&	0.08	&	12.06	&	30.39	&	0.46	&	0.16	&	0.5-6	&	0.93	&	0.36	&	12.06	&	8.55	&	(He$^3$,xn)	\\
$^{75}$Se	&	75	&	34	&	0.49	&	0.12	&	20.09	&	40.34	&	0.62	&	0.03	&	0.5-1.5	&	0.79	&	0.20	&	8.03	&	8.60	&	(d,4n)	\\
$^{75}$Se	&	75	&	34	&	0.69	&	0.19	&	20.09	&	42.92	&	0.62	&	0.03	&	0.5-1.5	&	1.12	&	0.31	&	8.03	&	8.60	&	(p,3n)	\\
$^{76}$Se	&	76	&	34	&	0.43	&	0.1	&	31.24	&	51.49	&	0.48	&	0.23	&	0-3.5	&	0.90	&	0.49	&	11.15	&	9.51	&	(d,4n)	\\
\hline
\end{tabular}
\end{table*}

\subsubsection{$^{nat}$Ge$(^3He ,xn)^{73}$Se$^{m,g}$}

\begin{figure}
\centering\includegraphics*[scale=\ones ]{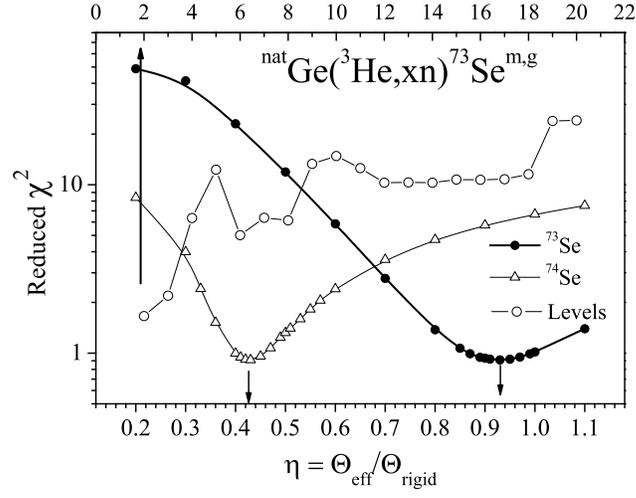}
\caption{\label{fig:he3-73Semg} Reduced $\chi ^2$ as a function  of the number of discrete levels and 
of the $\eta $ for $^{73}$Se and $^{74}$Se in the production of $^{73}$Se$^{m,g}$ in the $^{nat}$Ge$(He^3,xn)^{73}$Se$^{m,g}$ reaction. 
}
\end{figure}

\begin{figure}

\centering\includegraphics*[scale=\two ]{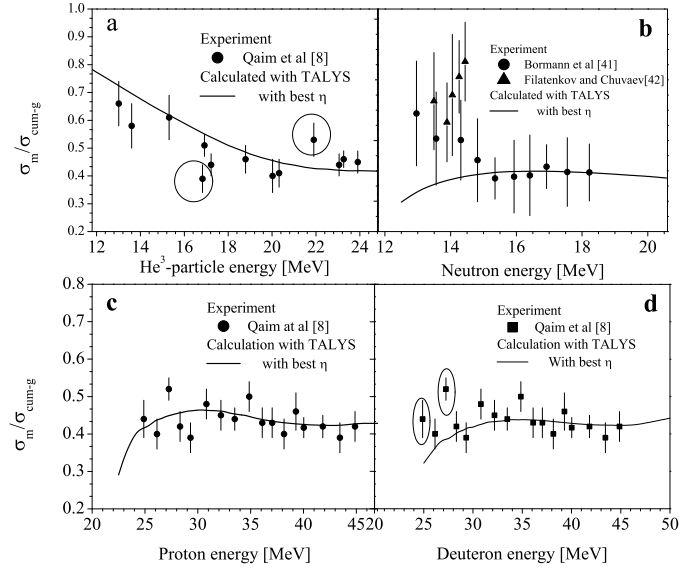}
\caption{\label{fig:xx73Se}Measured and calculated isomeric cross-section ratios for the $^{nat}$Ge$(^3He ,xn)^{73}$Se$^{m,g}$ (a),
 $^{74}$Se$(n,2n)^{73}$Se$^{m,g}$ (b), $^{75}$As$(p,3n)^{73}$Se$^{m,g}$ (c)
 and $^{75}$As$(d,4n)^{73}$Se$^{m,g}$ (d) 
reactions using the optimal number of discrete levels and the best $\eta$ value for each reaction. The encircled data points were removed from the fitting procedure.
}
\end{figure}

The calculation was carried out as described above. Mainly the $\eta $ of $^{73}$Se and $^{74}$Se have an effect on the isomeric cross-section ratio of  $^{73}$Se in the investigated incident energy range. (Some effect of $\eta $ on $^{75}$Se can be recognized, but its uncertainty is more than 100 p.c, therefore, it was omitted from further analysis.) 
Fig.~\ref{fig:he3-73Semg} depicts the reduced $\chi ^2$ as a function of the $\eta $ for $^{73}$Se and $^{74}$Se. The contribution of $^{70}$Ge, as target, to the isomeric cross section is negligible because the $^{73}$Se can be produced only  via  the $^{70}$Ge$(^3He , \gamma )^{73}$Se$^{m,g}$ reaction. 
The evaluated $\eta $ values for $^{73}$Se and $^{74}$Se are presented in Table ~\ref{tab:Se73-76}. Fig.~\ref{fig:xx73Se}a shows the measured and calculated isomeric cross-section ratio. 

\subsubsection{$^{74}$Se$(n,2n)^{73}$Se$^{m,g}$}

\begin{figure}
\centering\includegraphics*[scale=\ones ]{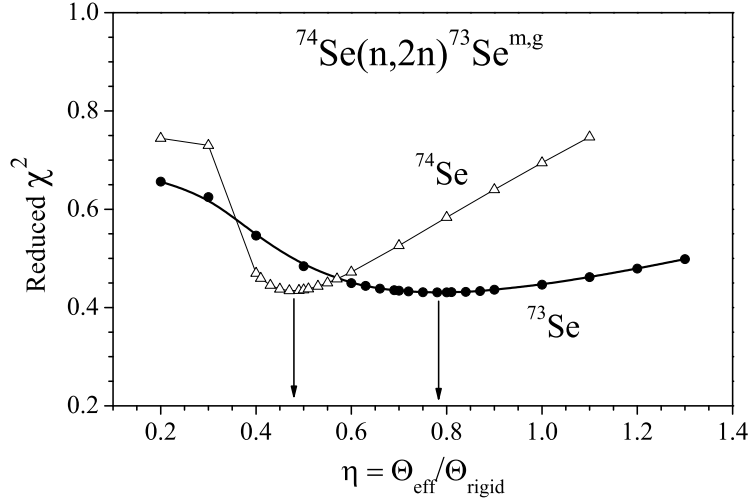}
\caption{\label{fig:n2n-73Semg} Reduced $\chi ^2$ as a function
of the $\eta $ for $^{73}$Se and $^{74}$Se in the production of $^{73}$Se$^{m,g}$ via the $^{74}$Se$(n,2n)^{73}$Se$^{m,g}$ reaction. 
}
\end{figure}

The experimental isomeric cross-section ratios were calculated from the cross section data of \citet{Bormann1976} and \citet{Filatenkov2001}. The calculation was carried out as described above. 
Fig.~\ref{fig:n2n-73Semg} displays the reduced $\chi ^2$ for this reaction having a quite weak minimum. 
The computed isomeric cross-section ratios depend on the $\eta $ values of $^{73}$Se and $^{74}$Se. 
Fig.~\ref{fig:xx73Se}b depicts the measured and calculated isomeric cross-section ratio. The evaluated $\eta $ values for $^{73}$Se and $^{74}$Se are presented in Table ~\ref{tab:Se73-76}.

\subsubsection{$^{75}$As$(p,3n)^{73}$Se$^{m,g}$}

\begin{figure}
\centering\includegraphics*[scale=\ones ]{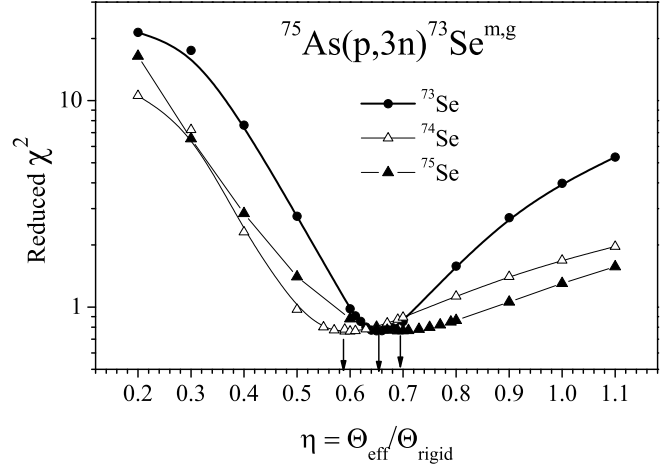}
\caption{\label{fig:p3n-73Semg} Reduced $\chi ^2$ as a function
of the $\eta $ for $^{73}$Se, $^{74}$Se and $^{75}$Se in the production of $^{73}$Se$^{m,g}$ via the $^{75}$As$(p,3n)^{73}$Se$^{m,g}$ reaction. 
}
\end{figure}

The reduced $\chi ^2$ for this reaction has a well defined minimum (Fig.~\ref{fig:p3n-73Semg}). The calculated isomeric cross-section ratios depend on the $\eta $ values of $^{73}$Se, $^{74}$Se and $^{75}$Se. 
The measured and calculated isomeric cross-section ratio is presented in Fig.~\ref{fig:xx73Se}c. The evaluated $\eta $ values for $^{73}$Se, $^{74}$Se and $^{75}$Se are given in Table ~\ref{tab:Se73-76}.

\subsubsection{$^{75}$As$(d,4n)^{73}$Se$^{m,g}$}

\begin{figure}
\centering\includegraphics*[scale=\ones ]{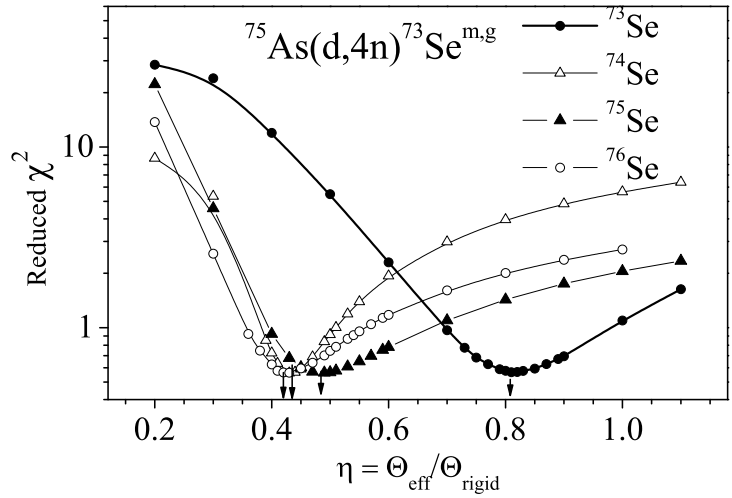}
\caption{\label{fig:d4n-73Semg} Reduced $\chi ^2$ as a function
of the $\eta $ for $^{73}$Se, $^{74}$Se $^{75}$Se and $^{76}$Se in the production of $^{73}$Se$^{m,g}$ via the $^{75}$As$(d,4n)^{73}$Se$^{m,g}$ reaction. 
}
\end{figure}

 The reduced $\chi ^2$  for this reaction has a well defined minimum as presented (Fig.~\ref{fig:d4n-73Semg}). The calculated isomeric cross-section ratios depend on the $\eta $ values of $^{73}$Se, $^{74}$Se, $^{75}$Se and $^{76}$Se. 
 The measured and calculated isomeric cross-section ratio is presented in Fig.~\ref{fig:xx73Se}d. The evaluated $\eta $ values for $^{73}$Se, $^{74}$Se, $^{75}$Se and $^{76}$Se are given in Table ~\ref{tab:Se73-76}.

They show a few interesting things: \newline
The $\eta $ value for $^{73}$Se was determined from four reactions, and the results agree well.  The simple average of the 4 data is  $\eta =0.79\pm 0.12$.  The weighted average (using the $1/(\Delta\eta)^2$ as weight factor) gives  $\eta =0.71\pm 0.06$.  
The data for the $^{74}$Se are in good agreement, the simple and weighted averages being $\eta =0.48\pm 0.08$ and $\eta =0.45\pm 0.05$,  respectively.

There are only two evaluations for $^{75}$Se which show quite good agreement within their uncertainties. Their simple and weighted averages are $\eta =0.59\pm 0.14$ and $\eta =0.55\pm 0.10$, respectively.

The evaluated $\eta $ values for $^{73}$Se, $^{74}$Se and $^{75}$Se confirm that the evaluation is independent of the reactions producing the isomeric state.

\subsection{$^{85}$Rb$(\alpha ,2n)^{87}$Y$^{m,g}$}

\begin{figure} 
\centering\includegraphics*[scale=\ones ]{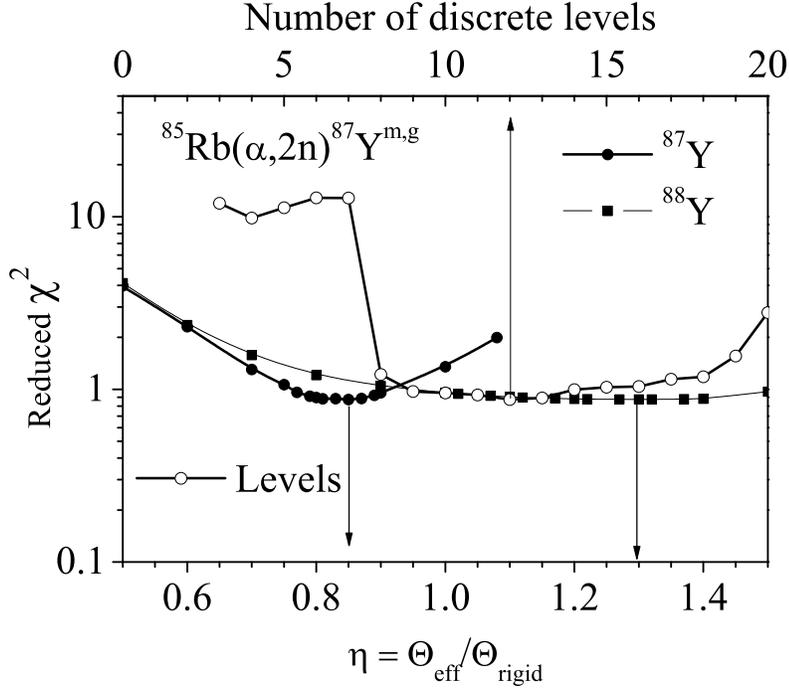}
\caption{\label{fig:87Y} Reduced $\chi ^2$ as a function of the number of discrete levels and $\eta $ of $^{87}$Y and $^{88}$Y  for the $^{85}$Rb$(\alpha ,2n)^{87}$Y$^{m,g}$ reaction. 
}
\end{figure}

\begin{figure}
\centering\includegraphics*[scale=\ones ]{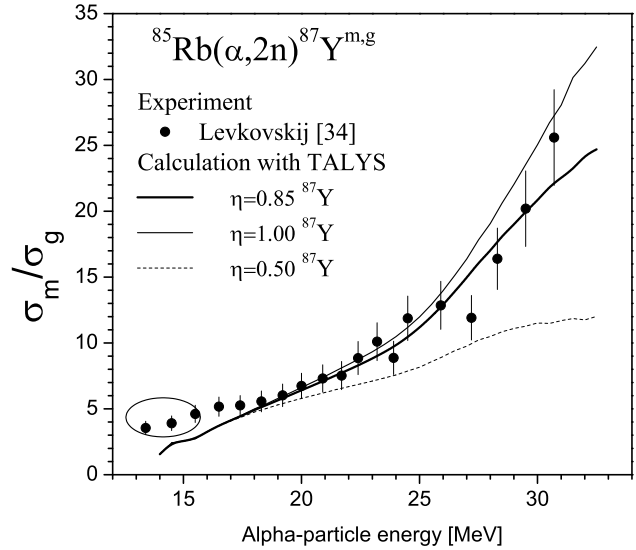}
\caption{\label{fig:87Ymg} Measured and calculated isomeric cross-section ratios for the $^{85}$Rb$(\alpha ,2n)^{87}$Y$^{m,g}$ reaction using the optimal number of discrete levels and the $\eta =0.85$, as well as two other value (1.00 and 0.30) for the $^{87}$Y. 
}
\end{figure}

\begin{table*}
\centering
\caption{\label{tab:Rb85}Summary of the evaluated continuum and discrete $\eta $ from different reactions producing $^{87,88}$Y, $^{90,91,92}$Nb, $^{94,95,96}$Tc, $^{108,109,110,111}$In and ENSDF \cite{ensdf} database.}
\begin{tabular}{cccccccccccccccc}
\hline
&&&Cont.&&&&Disc.&\\
Nucleus & A & Z & $\eta $ & $\Delta \eta $ & E$_x$ low & E$_x$ max & $\eta _d$ & $\Delta \eta _d$ & E$_x$ & $\eta $/$\eta _d$ & $\Delta$($\eta $/$\eta _d$) & S$_n$ & S$_p$ &	Reaction\\
\hline
$^{87}$Y & 87 & 39 & 0.85 & 0.13 & 0 & 16.79 & 0.68 & 0.28 & 0-3.5 & 1.26 & 0.56 & 11.81 & 5.78 & ($\alpha $,2n) \\
$^{88}$Y & 88 & 39 & 1.3 & 0.45 & 9.35 & 26.14 & 0.84 & 0.19 & 0-1.5 & 1.54 & 0.64 & 9.35 & 6.71 & ($\alpha $,2n) \\
$^{90}$Nb & 90 & 41	& 0.89 & 0.06 & 0 & 25.23 & 1.21 & 0.67 & 0-1.5 & 0.74 & 0.41 & 10.108 & 5.075 & ($\alpha $,3n)  \\
$^{91}$Nb & 91 & 41 & 0.75 & 0.15 & 12.05 & 37.27 & 0.66 & 0.14 & 0-3.5 & 1.13 & 0.33 & 12.048 & 5.154 & ($\alpha $,3n)  \\
$^{92}$Nb & 92 & 41 & 0.57 & 0.06 & 19.93 & 45.16 & 0.61 & 0.64 & 0-1.5 & 0.93 & 0.97 & 7.886 & 5.846 &	($\alpha $,3n)		\\
$^{94}$Tc & 94 & 43 & 0.93 & 0.04 & 0 & 20.86 & 0.61 & 0.30 & 0-1.5 & 1.54 & 0.75 & 8.623 & 4.64 & ($\alpha $,3n)\footnotemark[1] \\
$^{95}$Tc & 95 & 43 & 0.86 & 0.07 & 0 & 26.71 & 0.73 & 0.20 & 0-1.5 & 1.17 & 0.34 & 9.934 & 4.896 & ($\alpha $,2n)\footnotemark[2] \\
$^{95}$Tc & 95 & 43 & 1.1 & 0.26 & 9.93 & 30.79 & 0.73 & 0.20 & 0-1.5 & 1.50 & 0.55 & 9.934 & 4.896 & ($\alpha $,3n)\footnotemark[1] \\
$^{96}$Tc & 96 & 43 & 0.37 & 0.04 & 7.87 & 34.58 & 0.54 & 0.27 & 0-1.5 & 0.54 & 0.07 & 7.872 & 5.399 & ($\alpha $,2n)\footnotemark[2] \\
$^{96}$Tc & 96 & 43 & 1.15 & 0.16 & 17.81 & 38.66 & 0.68 & 0.08 & 0-1.5 & 1.69 & 0.31 & 7.872 & 5.399 & ($\alpha $,3n)\footnotemark[1] \\ 
$^{108}$In & 108 & 49 & 0.55 & 0.45 & 0.00 & 38.11 & 0.80 & 0.06 & 0-1.5 & 0.69 & 0.57 & 8.627 & 4.419 & (p,xn) \\
$^{109}$In & 109 & 49 & 0.35 & 0.13 & 10.44 & 48.55 & 0.45 & 0.24 & 0-1.5 & 0.79 & 0.51 & 10.439 & 4.524 & (p,xn) \\
$^{110}$In & 110 & 49 & 1.15 & 0.2 & 0.00 & 32.64 & 0.68 & 0.05 & 0-1.5 & 1.70 & 0.32 & 8.054 & 5.255 & (p,2n) \\
$^{111}$In & 111 & 49 & 0.44 & 0.06 & 9.99 & 42.63 & 0.41 & 0.23 & 0-1.5 & 1.08 & 0.60 & 9.991 & 5.331 & (p,2n)\\ 
\hline
\end{tabular}
\begin{flushleft}

\footnotemark[1] $\eta $ value from $^{93}$Nb$(\alpha ,3n)^{94}$Tc reaction

\footnotemark[2] $\eta $ value from $^{93}$Nb$(\alpha ,2n)^{95}$Tc reaction

\end{flushleft}
\end{table*}

The experimental isomeric cross-section ratios were calculated from the cross section data of \citet{Levk91}. The uncertainties of the isomeric cross-section ratios have been computed by formula \ref{eq:erreq}. 
The branching ratios of the levels in $^{87}$Y in the TALYS library were examined, and the simple approximation was substituted by our approach. The correction was necessary only to the level 7 ($2.321$ MeV) and above. 
The other problem was that experimental data existed below the reaction threshold, possible due to an energy shift. Therefore the lower three data were removed from the fitting procedure. 
Fig.~\ref{fig:87Y} shows the reduced $\chi ^2$ for comparing the model-calculated data with the experimental data. The best fit to the experimental data as a function of the number of levels (N) was obtained at N=12 but in this case the reduced $\chi ^2$ function is quite flat in the range of 9 to 16. 
There are four levels of the selected ones that have an uncertain spin assignment, which may have an effect on the evaluation. Fig.~\ref{fig:87Y} also shows the reduced $\chi ^2$ as a function of the parameter $\eta$ calculated using the fixed number of levels. 
The Table~\ref{tab:Rb85} shows the best fit to the experimental data achieved for $^{87}$Y and $^{88}$Y. Fig.~\ref{fig:87Ymg} shows the measured and calculated isomeric cross-section ratios as a function of the $\alpha $-particle energy for the best $\eta$ value as well as for two other values.

%
%

\subsection{$^{89}$Y$(\alpha ,3n)^{90}$Nb$^{m,g}$}

\begin{figure}
\centering\includegraphics*[scale=\ones ]{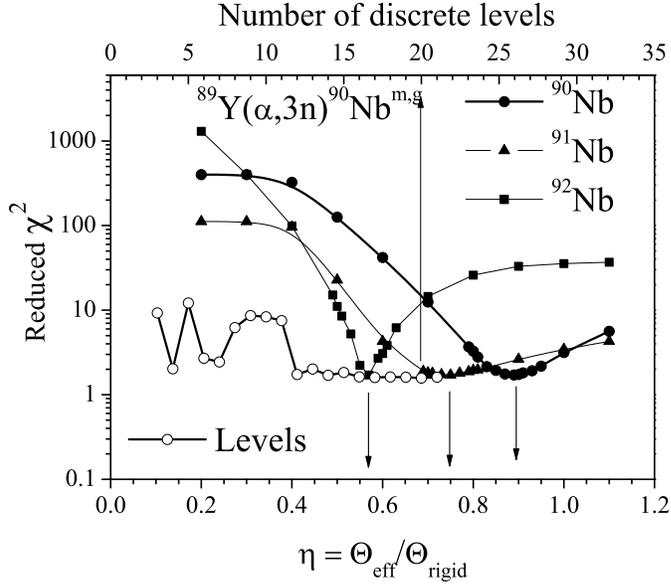}
\caption{\label{fig:90Nb} Reduced $\chi ^2$ as a function of the number of discrete levels and $\eta $ for the $^{89}$Y$(\alpha ,3n)^{90}$Nb$^{m,g}$ reaction. 
}
\end{figure}

\begin{figure}
\centering\includegraphics*[scale=\ones ]{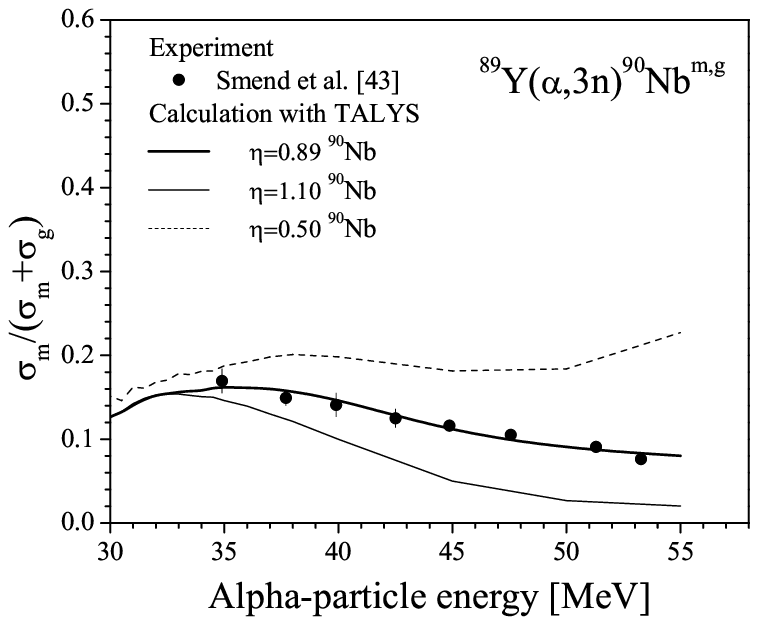}
\caption{\label{fig:90Nbmg} Measured and calculated isomeric cross-section ratios for the 
 $^{89}$Y$(\alpha ,3n)^{90}$Nb$^{m,g}$ reaction using the optimal number of discrete
 levels and the $\eta $ values of 0.89, 1.10 and 0.50. 
}
\end{figure}

The experimental isomeric cross-section ratios and their uncertainties were taken from \citet{smend67}. The data were reported as $\sigma _g /\sigma _m$ ratios, but we interpret them as $(\sigma _g +\sigma _m)/\sigma _m$ ratios, 
because the isomer decays fully into the ground state and it seems logical that $\sigma _g $ refers to $(\sigma _g +\sigma _m)$. 
The second reason is that the fitting procedure gives a far better reduced $\chi ^2$ for this interpretation. (The $\eta $ value belonging to the minimal reduced $\chi ^2$ would have been a bit higher under the original assumption.) 
Fig.~\ref{fig:90Nb} depicts the reduced $\chi ^2$ for comparing the model-calculated data and the experimental data. The best fit to the experimental data as a function of the number of levels (N) was obtained at N=20. 
The figure shows that after the first oscillations the fit becomes better and better with the increasing number of levels, and it becomes almost constant in the range of 15-21; after that, it increases again. 
Fig.~\ref{fig:90Nb} also displays the reduced $\chi ^2$ as a function of the $\eta$ of the $^{90}$Nb, $^{91}$Nb and $^{92}$Nb nuclei.
Table~\ref{tab:Rb85} presents the best fit $\eta$ values achieved for the $^{90}$Nb, $^{91}$Nb and $^{92}$Nb isotopes by comparison of the measured and calculated isomeric cross-section ratios. 
Fig.~\ref{fig:90Nbmg} depicts the measured and calculated isomeric-cross section ratios as a function of the incident $\alpha $-particle energy for the best fit with optimal $\eta $ values as well as fits with $\eta $ values of 1.10 and 0.50 of $^{90}$Nb.


\subsection{$^{93}$Nb$(\alpha ,2n)^{95}$Tc$^{m,g}$}

\begin{figure}%
\centering\includegraphics*[scale=\ones ]{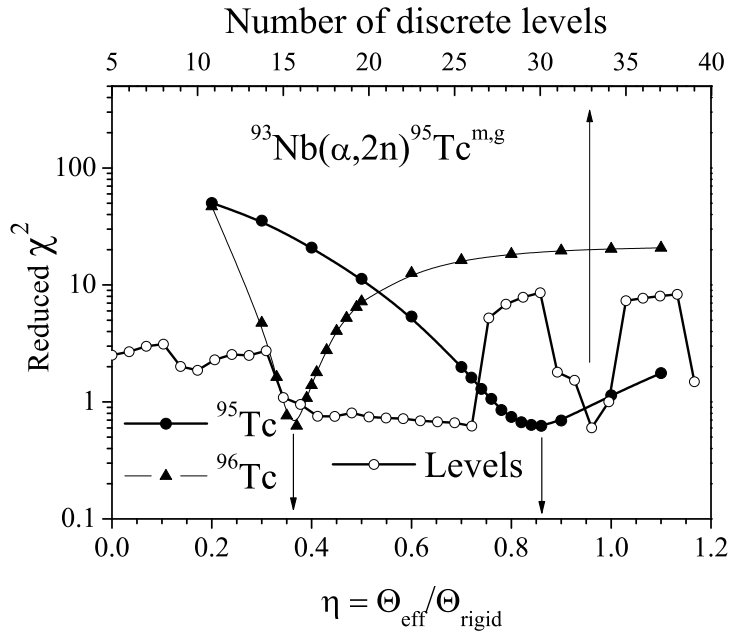}
\caption{\label{fig:95Tc} Reduced $\chi ^2$ as a function of the number of discrete levels and $\eta $ of the $^{95}$Tc and $^{96}$Tc nuclei in the $^{93}$Nb$(\alpha ,2n)^{95}$Tc$^{m,g}$ reaction. 
}
\end{figure}

\begin{figure}
\centering\includegraphics*[scale=\ones ]{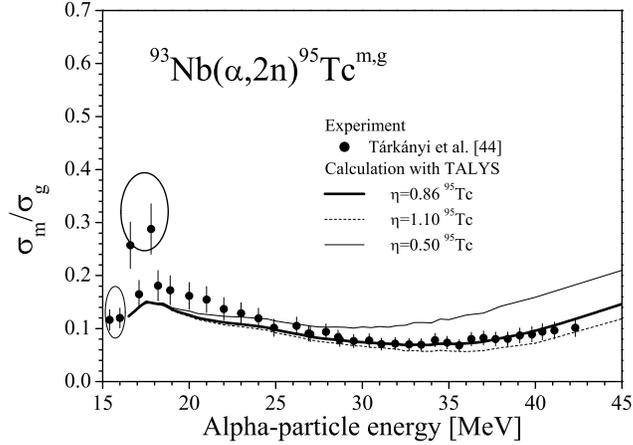}
\caption{\label{fig:95Tcmg} Measured and calculated isomeric cross-section ratios for the 
 $^{93}$Nb$(\alpha ,2n)^{95}$Tc$^{m,g}$ reaction using the optimal number of discrete levels
 and the $\eta $ values of 0.86, 1.10 and 0.50 of $^{95}$Tc. 
The four points removed are encircled. 
}
\end{figure}

The experimental isomeric cross-section ratios were calculated from the cross section data of \citet{Tarkanyi2002}. The uncertainties of the isomeric cross-section ratios have been computed by formula \ref{eq:erreq}. Fig.~\ref{fig:95Tc} depicts the reduced $\chi ^2$ for comparing the model-calculated data with the experimental data. 
The best fit to the experimental data as a function of the number of levels (N) was obtained at N=33. In contrast to $^{90}$Nb, the reduced $\chi ^2$ here was not stabilized with the increasing number of the discrete levels used. 
The best fit of the measured and calculated isomeric cross-section ratios can be achieved by $\eta =0.86\pm 0.07$ and $\eta =0.37\pm 0.04$ for the $^{95}$Tc and $^{96}$Tc nuclei, respectively. 
The data for $^{96}$Tc nucleus are in strong disagreement with the evaluation result from the $^{93}$Nb$(\alpha ,3n)^{94}$Tc$^{m,g}$ data and the systematics; therefore it was not used in the evaluation. 
The relevant data are summarized in Table~\ref{tab:Rb85}. Fig.~\ref{fig:95Tcmg} depicts the measured and calculated isomeric cross-section ratios as a function of the incident $\alpha $ particle energy for the best fit with $\eta =0.86$ as well as two other fits with($\eta =1.10$, $\eta =0.50$) of $^{95}$Tc. 
The deviation of the measured from the calculated isomeric cross-section ratio shows a systematic trend as a function of the incident $\alpha $ particle energy: 
The measured data are higher than the calculated ones below 25 MeV, and the calculated data are almost independent of the $\eta $ parameter. 
This indicates that the level scheme is not correct and this causes the problem with the $\eta $ value of  $^{96}$Tc. There is an experimental problem with this measurements too: data are presented below the threshold of the reaction.

\subsection{$^{93}$Nb$(\alpha ,3n)^{94}$Tc$^{m,g}$}

\begin{figure}
\centering\includegraphics*[scale=\ones ]{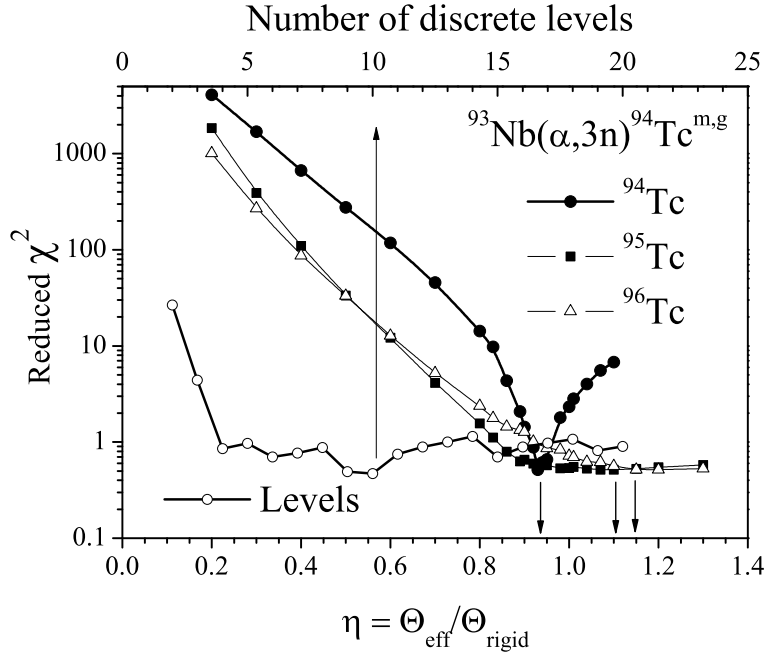}
\caption{\label{fig:94Tc} Reduced $\chi ^2$ as a function of the number of discrete levels 
and $\eta $ of the $^{94}$Tc, $^{95}$Tc and $^{96}$Tc nuclei in the $^{93}$Nb$(\alpha ,3n)^{94}$Tc$^{m,g}$ reaction. 
}
\end{figure}

\begin{figure}
\centering\includegraphics*[scale=\ones ]{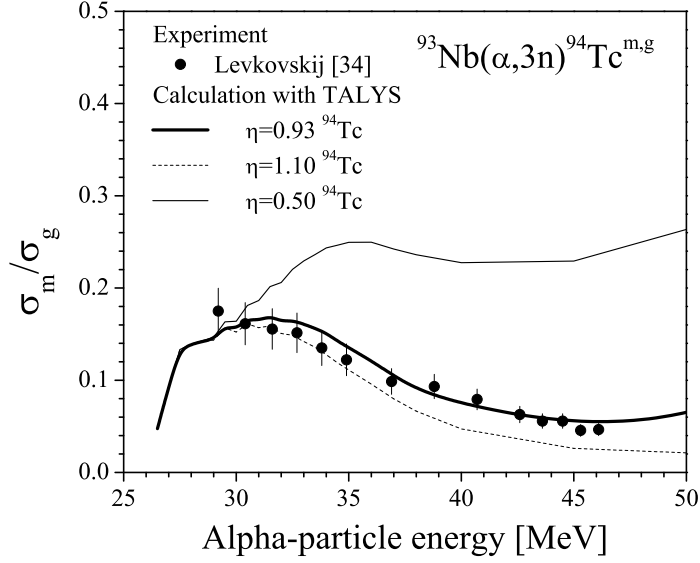}
\caption{\label{fig:94Tcmg} Measured and calculated isomeric cross-section ratios for the $^{93}$Nb$(\alpha ,3n)^{94}$Tc$^{m,g}$ reaction using the optimal number of discrete levels and the $\eta =0.93$, $\eta =1.10$ and $\eta =0.50$ of $^{94}$Tc. 
}
\end{figure}

The experimental isomeric cross-section ratios were calculated from the cross section data of \citet{Levk91}.  
The formula \ref{eq:erreq} was used to calculate their uncertainties. The branching ratios of the levels of $^{94}Tc$ in the TALYS library were known except for the level 2 ($98 $ keV), which can decay to the ground and the metastable state. 
This branching ratio has a very strong effect on the isomeric cross-section ratio. We found the best value of 0.40 to the ground and 0.60 to the metastable state. 
The achieved minimum values of the reduced $\chi ^2$ reach the statistically required value near one, as seen in Fig.~\ref{fig:94Tc}, which shows the reduced $\chi ^2$ for comparing the model-calculated data with the experimental data. The best fit to the experimental data as a function of the number of levels (N) was obtained at N=10. 
Fig.~\ref{fig:94Tc} also shows the reduced $\chi ^2$ as a function of the parameter $\eta$ calculated after fixing the number of levels. 
The best fit to the experimental data are summarized in Table~\ref{tab:Rb85} for the $^{94}$Tc, $^{95}$Tc and $^{96}$Tc nuclei. 
 Fig.~\ref{fig:94Tcmg} depicts the measured and calculated isomeric-cross section ratios as a function of the incident $\alpha $ particle energy for the best fit with $\eta =0.93$ and two other values ($\eta =1.10$, $\eta =0.50$) of $^{94}$Tc.
 
 The two reactions above give the possibility to compare the $\eta$ values of $^{95}$Tc and $^{96}$Tc nuclei. In the case of  $^{95}$Tc the $\eta$ values are in agreement within the estimated uncertainties, but for the $^{96}$Tc nucleus, the $\eta$ values are strongly incompatible.


-
\subsection{$^{111}$Cd$(p,2n)^{110}$In$^{m,g}$}

\begin{figure}
\centering\includegraphics*[scale=\ones ]{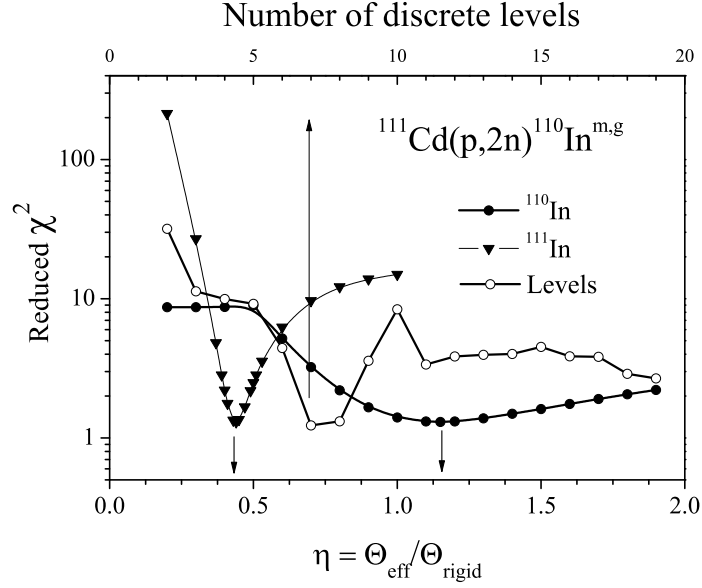}
\caption{\label{fig:110In} Reduced $\chi ^2$ as a function of the number of discrete levels and $\eta $ of the $^{110}$In and $^{111}$In nuclei in the $^{111}$Cd$(p,2n)^{110}$In$^{m,g}$ reaction. 
}
\end{figure}

\begin{figure}
\centering\includegraphics*[scale=\ones ]{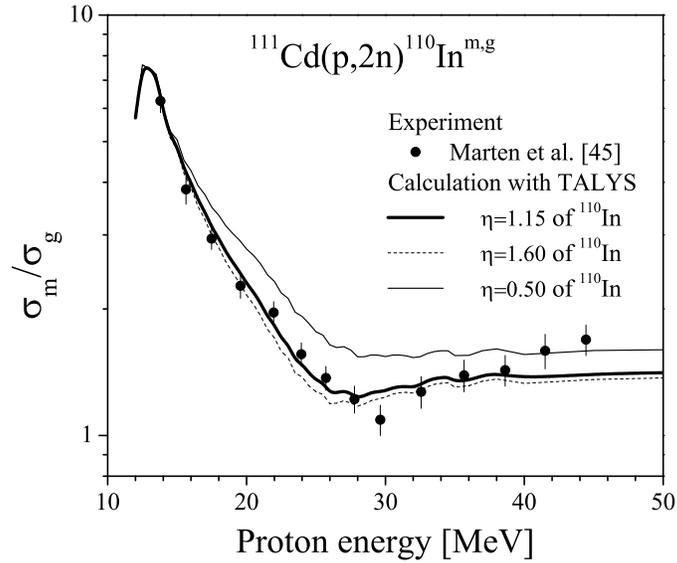}
\caption{\label{fig:110Inmg} Measured and calculated isomeric cross-section ratios for the $^{111}$Cd$(p,xn)^{110}$In$^{m,g}$ reaction using the optimal number of discrete levels and $\eta $ values of 1.15, 1.16 and 0.50 for $^{110}$In and the optimal value of $^{111}$In. 
}
\end{figure}

The experimental isomeric cross-section ratios were obtained from the data of \citet{Marten1985}. Fig.~\ref{fig:110In} depicts the reduced $\chi ^2$ for comparing the model-calculated data with the experimental data. 
The best fit to the experimental data as a function of the number of levels (N) was obtained at N=7. Table~\ref{tab:Rb85} summarizes the best fit $\eta$ values for the $^{110}$In and $^{111}$In nuclei. 
Fig.~\ref{fig:110Inmg} shows the measured and calculated isomeric cross-section ratios using the optimal number of discrete levels and for the best fit with $\eta =1.15$ and two other values ($\eta =1.60$, $\eta =0.50$) of $^{110}$In.

\subsection{$^{nat}$Cd$(p,xn)^{108}$In$^{m,g}$}

\begin{figure}
\centering\includegraphics*[scale=\ones ]{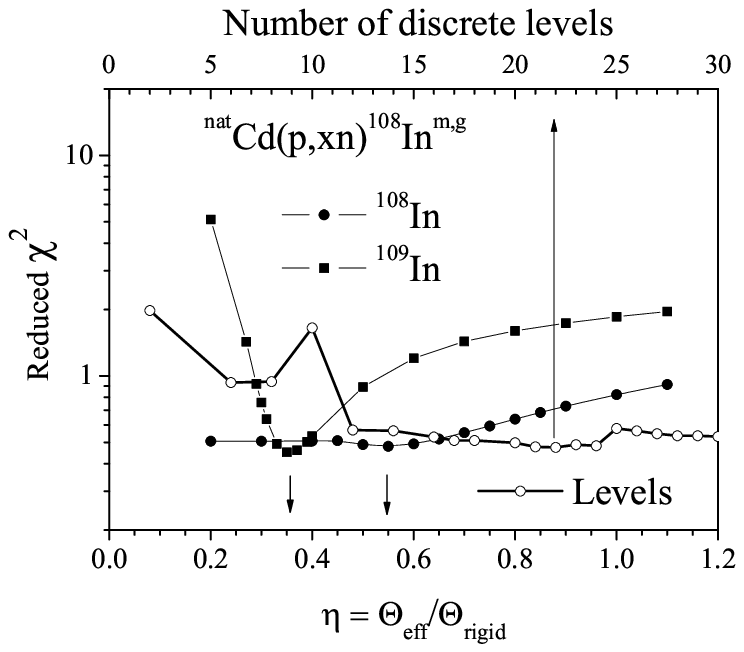}
\caption{\label{fig:108In} Reduced $\chi ^2$ as a function of the number of discrete levels and $\eta $ for the $^{108}$In and $^{109}$In nuclei in the $^{nat}$Cd$(p,xn)^{108}$In$^{m,g}$ reaction. 
}
\end{figure}

\begin{figure}
\centering\includegraphics*[scale=\ones ]{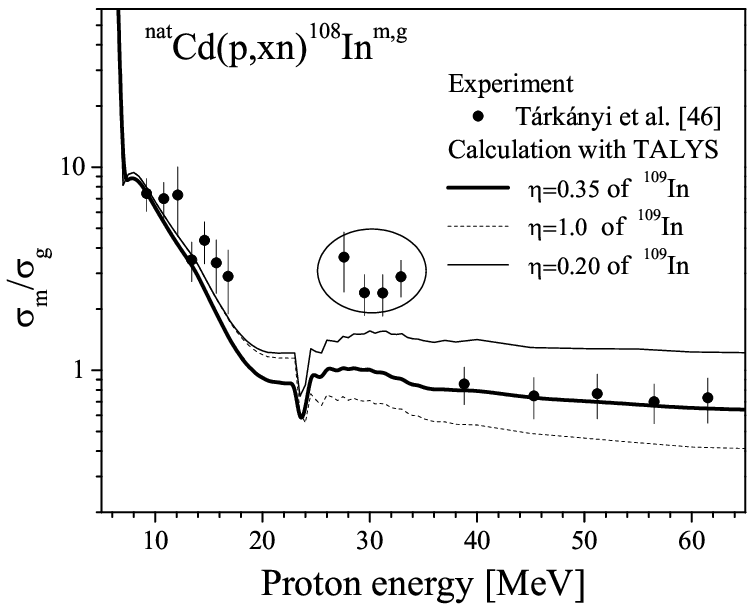}
\caption{\label{fig:108Inmg} Measured and calculated isomeric cross-section ratios for the $^{nat}$Cd$(p,xn)^{108}$In$^{m,g}$ reaction using the optimal number of discrete levels and for the best fit with $\eta =0.35$ and two other values ($\eta =1.0$, $\eta =0.20$) of $^{109}$In and the optimal value of $^{108}$In. 
 The points removed are encircled.
}
\end{figure}

The experimental isomeric cross-section ratios were calculated from the cross section data of \citet{Tarkanyi2006}. The uncertainties of the isomeric cross-section ratios were calculated as above. 
Using the cross section measured on an element of  natural isotopic composition is not ideal for the comparison of experimental and model calculated isomeric cross-section ratio because canceling the effect of the optical model parameters is not valid. 
The second problem in this case is that too many parameters would need to be fited. Therefore $\eta$ values of the $^{110}$In and $^{111}$In nuclei were fixed to the values determined from the $^{111}$Cd$(p,2n)^{110}$In$^{m,g}$ reaction. 
Fig.~\ref{fig:108In} depicts the reduced $\chi ^2$ for comparing the model-calculated data and the experimental data. The $\chi ^2$ as a function of the number of levels (N) has a minimum at the value of N equal to 22. 
The $\chi ^2$ value as a function of the $\eta$ value of $^{108}$In has only a feeble minimum. The best estimation of $\eta$ values for the $^{108}$In and $^{109}$In nuclei are presented in Table~\ref{tab:Rb85}. 
Fig.~\ref{fig:108Inmg} depicts the measured and calculated isomeric-cross section ratios as a function of the incident proton energy for the best fit with $\eta =0.35$ and two other values ($\eta =1.0$, $\eta =0.20$) of $^{109}$In. 
Some experimental points were removed from the fit because of their large deviation from the calculation and the general trend of the experimental data.

\begin{table*}
\centering
\caption{\label{tab:Te127}Summary of the evaluated continuum and discrete $\eta $ from different reactions producing $^{121,122,123,127}$Te, $^{139,140,141,142}$Nd, $^{159}$Tb, $^{183,184}$Re, $^{195,196,197,198}$Hg, $^{198,199,200}$Tl and ENSDF \cite{ensdf} database.}
\begin{tabular}{cccccccccccccccc}
\hline
&&&Cont.&&&&Disc.&\\
Nucleus & A & Z & $\eta $ & $\Delta \eta $ & E$_x$ low & E$_x$ max & $\eta _d$ & $\Delta \eta _d$ & E$_x$ & $\eta $/$\eta _d$ & $\Delta$($\eta $/$\eta _d$) & S$_n$ & S$_p$ &	Reaction\\
\hline
$^{121}Te$ & 121 & 52 & 0.55 & 0.17 & 0.00 & 28.88 & 0.44 & 0.09 & 0-2 & 1.25 & 0.47 & 7.25 & 7.42 & ($\alpha $,xn)\\
$^{122}Te$ & 122 & 52 & 0.17 & 0.03 & 9.84 & 38.72 & 0.26 & 0.08 & 0-3 & 0.65 & 0.22 & 9.84 & 8.00 & ($\alpha $,xn)\\
$^{123}Te$ & 123 & 52 & 0.33 & 0.11 & 16.77 & 45.64 & 0.45 & 0.20 & 0-2.5 & 0.73 & 0.40 & 6.93 & 8.13 & ($\alpha $,xn)\\ 
$^{122}Sb$ & 122 & 51 & 0.77 & 0.09 & 0 & 6.36 & 0.46 & 0.24 & 0-0.5 & 0.72 & 0.36 & 6.81 & 6.43 & (p,n) \\
$^{141}Nd$ & 141 & 60 & 0.2 & 0.15 & 0.00 & 15.35 & 0.39 & 0.22 & 0-3 & 0.51 & 0.48 & 8.01 & 6.79 & ($^3He $,xn) \\
$^{142}Nd$ & 142 & 60 & 0.15 & 0.15 & 9.83 & 25.18 & 0.27 & 0.12 & 2-4.5 & 0.56 & 0.62 & 9.83 & 7.22 & ($^3He $,xn) \\
$^{141}Nd$ & 141 & 60 & 0.2 & 0.06 & 0.00 & 16.10 & 0.39 & 0.22 & 0-3 & 0.51 & 0.29 & 8.01 & 6.79 & (p,n) \\
$^{139}Nd$ & 139 & 60 & 0.56 & 0.09 & 0.00 & 42.39 & 0.69 & 0.24 & 0-2.5 & 0.81 & 0.30 & 8.01 & 6.79 & (p,3n) \\
$^{140}Nd$ & 140 & 60 & 0.24 & 0.06 & 10.31 & 52.70 & 0.21 & 0.11 & 0-3 & 1.14 & 0.58 & 10.31 & 6.72 & (p,3n) \\
$^{141}Nd$ & 141 & 60 & 0.65 & 0.09 & 18.32 & 60.71 & 0.39 & 0.22 & 0-3 & 1.66 & 0.95 & 8.01 & 6.79 & (p,3n) \\
$^{162}Ho$ & 162 & 67 & 0.43 & 0.06 & 0.00 & 17.52 & 0.84 & 0.37 & 0-1 & 0.51 & 0.23 & 6.916 & 5.275 & ($\alpha $,n) \\
$^{183}Re$ & 183 & 75 & 0.25 & 0.03 & 8.43 & 20.14 & 0.36 & 0.04 & 0-1 & 0.70 & 0.12 & 8.43 & 4.85 & ($\alpha $,3n) \\
$^{184}Re$ & 184 & 75 & 0.36 & 0.07 & 14.92 & 26.62 & 0.31 & 0.03 & 0-1 & 1.16 & 0.24 & 6.49 & 5.16 & ($\alpha $,3n) \\
$^{195}Hg$ & 195 & 80 & 0.21 & 0.04 & 0.00 & 29.82 & 0.27 & 0.10 & 0-0.5 & 0.79 & 0.33 & 6.89 & 6.08 & ($^3He $,xn) \\
$^{196}Hg$ & 196 & 80 & 0.19 & 0.03 & 8.90 & 38.72 & 0.11 & 0.07 & 0-2 & 1.65 & 1.11 & 8.90 & 6.55 & ($^3He $,xn) \\
$^{197}Hg$ & 197 & 80 & 0.2 & 0.04 & 15.69 & 45.51 & 0.28 & 0.02 & 0-0.5 & 0.72 & 0.17 & 6.79 & 6.69 & ($^3He $,xn) \\
$^{198}Hg$ & 198 & 80 & 0.25 & 0.17 & 22.58 & 52.40 & 0.17 & 0.09 & 0-0.5 & 1.49 & 1.30 & 6.89 & 6.08 & ($^3He $,xn) \\
$^{197}Hg$ & 197 & 80 & 0.2 & 0.10 & 0.00 & 10.18 & 0.28 & 0.02 & 0-0.5 & 0.72 & 0.36 & 6.79 & 6.69 & ($\alpha $,xn) \\
$^{198}Hg$ & 198 & 80 & 0.19 & 0.10 & 6.89 & 17.07 & 0.17 & 0.09 & 0-0.5 & 1.14 & 0.85 & 6.89 & 6.08 & ($\alpha $,xn) \\
$^{198}Tl$ & 198 & 81 & 0.25 & 0.05 & 0.00 & 25.49 & 0.33 & 0.38 & 0-1 & 0.75 & 0.87 & 7.23 & 4.24 & ($\alpha $,3n) \\
$^{199}Tl$ & 199 & 81 & 0.10 & -- & 8.64 & 34.13 & 0.15 & 0.12 & 0-1 & --& -- & 8.64 & 4.39 & ($\alpha $,3n) \\
$^{200}Tl$ & 200 & 81 & 0.23 & 0.02 & 15.70 & 41.19 & 0.16 & 0.18 & 0-1 & 1.40 & 1.53 & 7.06 & 4.79 & ($\alpha $,3n)\\
\hline
\end{tabular}
\begin{flushleft}

\end{flushleft}
\end{table*}


\subsection{$^{nat}$Sn$(\alpha ,xn)^{121}$Te$^{m,g}$}

\begin{figure}
\centering\includegraphics*[scale=\ones ]{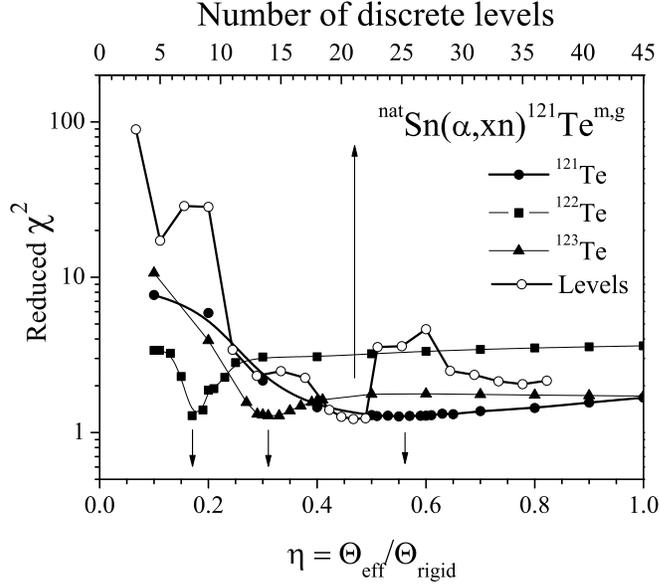}
\caption{\label{fig:121Te} Reduced $\chi ^2$ as a function of the number of discrete levels and $\eta $ for the $^{121}$Te, $^{122}$Te and $^{123}$Te in the $^{nat}$Sn$(\alpha ,xn)^{121}$Te$^{m,g}$ reaction. 
}
\end{figure}

\begin{figure}
\centering\includegraphics*[scale=\ones ]{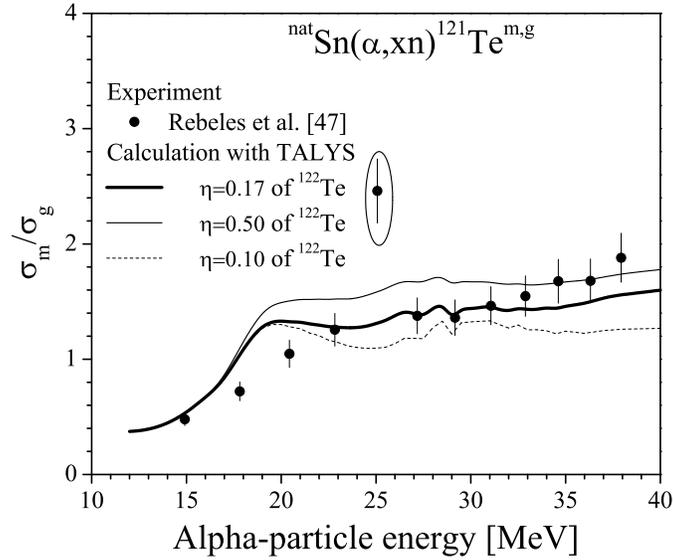}
\caption{\label{fig:121Temg} Measured and calculated isomeric cross-section ratios for the $^{nat}$Sn$(\alpha ,xn)^{121}$Te$^{m,g}$ reaction using the optimal number of discrete levels and for the best fit with $\eta =0.17$ and two other values ($\eta =0.1$, $\eta =0.50$) of $^{122}$Te and the optimal value of $^{121}$Te and $^{123}$Te. The point removed from the fit is encircled.
}
\end{figure}

The experimental isomeric cross-section ratios were calculated from the cross section data of \citet{Rebeles2007}. The uncertainties of the isomeric cross section ratios were calculated as above. 
Fig.~\ref{fig:121Te} depicts the reduced $\chi ^2$ for comparing the model-calculated data with the experimental data. The best fit to the experimental data as a function of the number of levels (N) was obtained at N=21. 
The figure shows that the proper selection of the discrete level is crucial. 
The shape of the reduced $\chi ^2$ curve is different for the three nuclei considered. 
The sharpest one is for the $^{122}$Te and the least significant minimum occurs for the $^{121}$Te. 
The best estimation of $\eta$ values for the $^{121}$Te, $^{122}$Te and $^{123}$Te nuclei and their comparison with the $\eta_d$ values derived from the discrete levels are summarized in Table~\ref{tab:Te127}. 
Fig.~\ref{fig:121Temg} shows the measured and calculated isomeric-cross section ratios as a function of the incident $\alpha $-particle energy for the best fit with the $\eta =0.17$ and two other values ($\eta =0.1$, $\eta =0.50$) of $^{122}$Te and the optimal value of $^{121}$Te and $^{123}$Te were fixed.


\subsection{$^{122}$Sn$(p,n)^{122}$Sb$^{m,g}$}

\begin{figure}
\centering\includegraphics*[scale=\ones ]{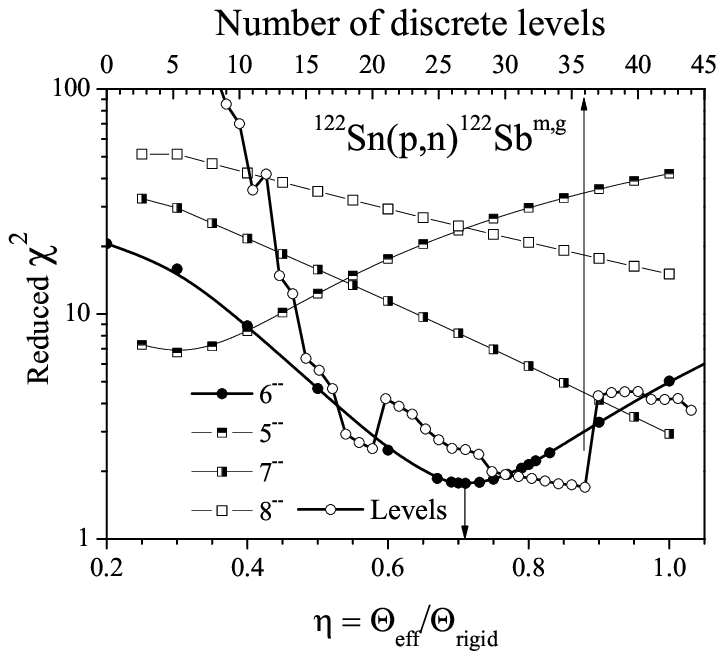}
\caption{\label{fig:122Sb} Reduced $\chi ^2$ as a function of the number of discrete levels and $\eta $ for the $^{122}$Sn$(p,n)^{122}$Sb$^{m,g}$ reaction. Calculations were carried out using the values  $8^-, 7^-, 6^-$ and $5^-$ for spin and parity of the isomeric state.
}
\end{figure}

\begin{figure}
\centering\includegraphics*[scale=\ones ]{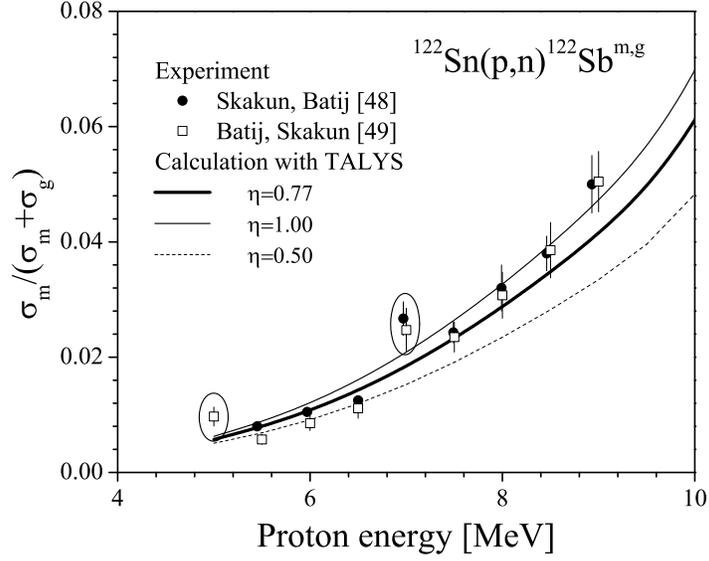}
\caption{\label{fig:122Sbmg} Measured and calculated isomeric cross-section ratios for the $^{122}$Sn$(p,n)^{122}$Sb$^{m,g}$ reaction using the optimal number of discrete levels and the $\eta =0.77$, $\eta =0.50$ and $\eta =1.0$ . The points removed are encircled.
}
\end{figure}

The experimental isomeric cross-section ratios were calculated from the cross-section data of 
 \citet{skakun1992}
and \citet{Batij2007}. In the first paper  the isomeric cross-section ratios were published. In the second one, the ratios were calculated from the presented cross-section data. 
While the authors are the same and isomeric cross-section ratios are very similar, they seem to be strongly correlated. 
The branching ratios of the levels of $^{122}$Sb in the TALYS library were examined, and the simple approximation was substituted by our approach. The ENSDF \cite{ensdf} library data for $^{122}$Sb were checked, and it came to light, that the spin of the isomeric level (8) is only a suggested value. 
There is no strong experimental evidence supporting the value 8. In this work, at first calculations were carried out with 42 discrete levels including the default spin value 8. 
Over the whole investigated energy range, the calculated isomeric cross-section ratio was similar to the measured one, but the magnitude of the calculated one was always lower than the measured one using any $\eta $ value. Fig.~\ref{fig:122Sb} depicts  reduced $\chi ^2$. 
The curve obtained using a spin value of $8^-$decreases monotonously without a minimum and is very far from the statistically acceptable value of 1. The calculation was repeated with $7^-$, $6^-$ and $5^-$ values of the spin of the isomeric level and Fig.~\ref{fig:122Sb} shows those results too. 
It is evident that in the case of a spin value of $6^-$ the reduced $\chi ^2$ curve shows an acceptable behavior. It has a minimum at $\eta =0.77$, and the reduced $\chi ^2$ is quite near to 1. 
It may be important to remark that the curve with $5^-$ shows an opposite behavior compared to the others, increasing with $\eta $ and having a minimum at a low $\eta $ value, but the minimum is far from the fair value of one. 
These curves indicate that the spin assignment of the isomeric level is critical. Before the final evaluation of the reduced $\chi ^2$ curve was done, the dependence on the number of discrete levels used was reevaluated as well as the TALYS branching data recalculated using the spin value of $6^-$ for the isomeric level. 
The best fit to the experimental data as a function of the number of levels (N) was obtained at N=36. 
The best fit to the experimental data can be achieved by using $\eta =0.77\pm 0.09$ (see Fig.~\ref{fig:122Sbmg}). The Table~\ref{tab:Te127} presents the best estimation of $\eta$ value and comparison with the $\eta$ value derived from the discrete levels.

\subsection{$^{nat}$Ce$(^3He,xn)^{139,141}$Nd$^{m,g}$ and $^{141}$Pr$(p,xn)^{139,141}$Nd$^{m,g}$ }

\begin{figure}
\centering\includegraphics*[scale=\ones ]{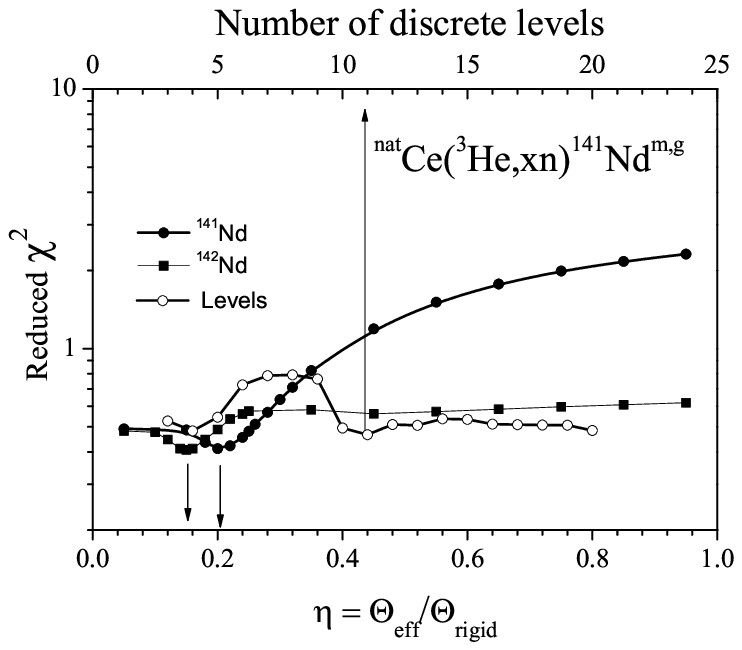}
\caption{\label{fig:141Nd-Ce} Reduced $\chi ^2$ as a function of the number of discrete levels and $\eta $ for the $^{nat}$Ce$(^3He,xn)^{141}$Nd$^{m,g}$ reaction. 
}
\end{figure}

\begin{figure}
\centering\includegraphics*[scale=\ones ]{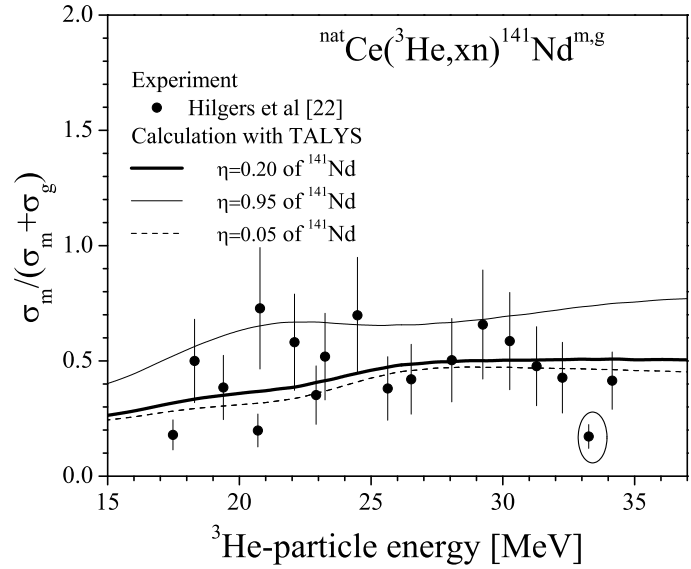}
\caption{\label{fig:141Ndmg-Ce} Measured and calculated isomeric cross-section ratios for the $^{nat}$Ce$(^3He,xn)^{141}$Nd$^{m,g}$ reaction using the optimal number of discrete levels and the $\eta =0.20$, $\eta =0.05$ and $\eta =0.95$ for the $^{141}$Nd. An experimental point removed from the fit is encircled.
}
\end{figure}

\begin{figure}
\centering\includegraphics*[scale=\ones ]{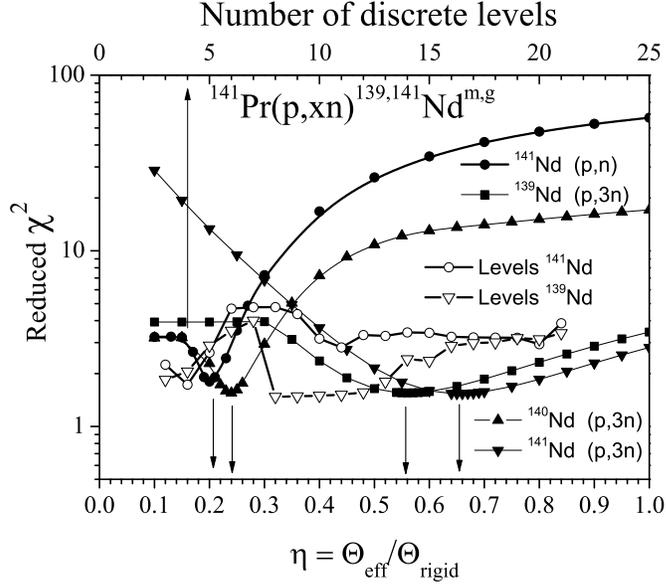}
\caption{\label{fig:141Nd-Pr} Reduced $\chi ^2$ as a function of the number of discrete levels and $\eta $ for the $^{141}$Pr$(p,xn)^{139,141}$Nd$^{m,g}$ reaction. 
}
\end{figure}

\begin{figure}
\centering\includegraphics*[scale=\ones ]{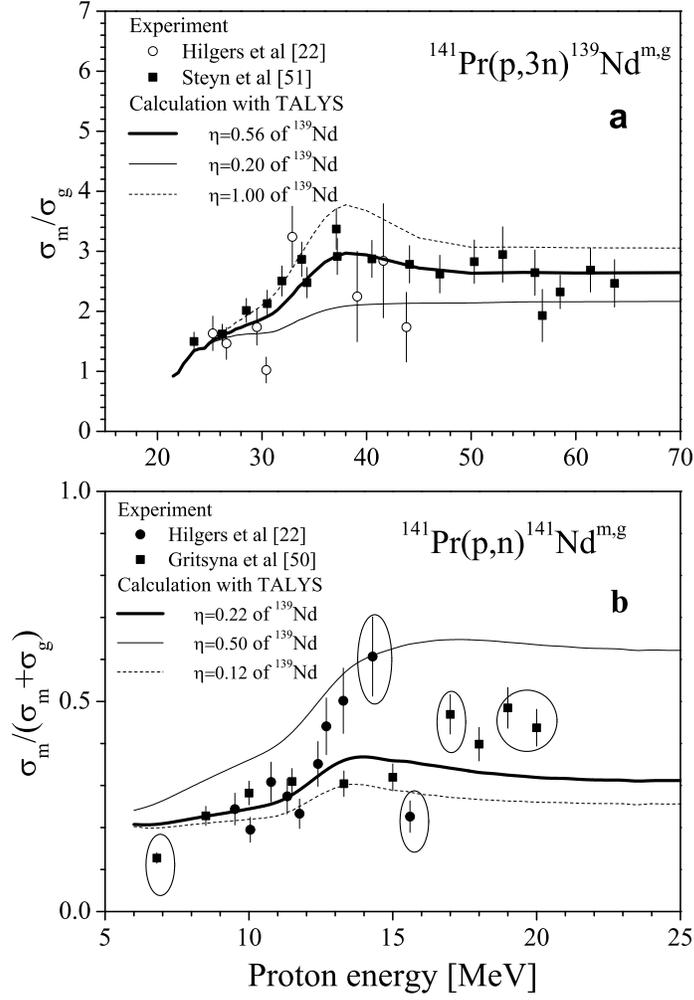}
\caption{\label{fig:141Ndmg-Pr} Measured and calculated isomeric cross-section ratios for the $^{141}$Pr$(p,xn)^{139,141}$Nd$^{m,g}$ reaction using the optimal number of discrete levels, $\eta $ values for the best fit and two other values. The experimental points removed from the fit are encircled.
}
\end{figure}

Calculations were done by us earlier \cite{Hilgers2007} using the STAPRE nuclear reaction model code. Now the experimental data were reevaluated using the method described here. 
The experimental data from \citet{Hilgers2007} were used for the $^3He$ and $p$ induced reactions while \citet{Gritsyna1963} and \citet{Steyn2006} data were used only for the $p$ induced reactions. 
Fig.~\ref{fig:141Nd-Ce} shows the reduced $\chi ^2$ for comparing the model-calculated data and the experimental data for the $^{141}$Nd as a function of $\eta $ values of the $^{141}$Nd and $^{142}$Nd and number of the discrete levels of $^{141}$Nd. 
The best fit to the experimental data as a function of the number of levels (N) was obtained at N=11. The best estimation of $\eta$ values for the $^{141}$Nd and $^{142}$Nd nuclei and their comparison with the $\eta$ values derived from the discrete levels are summarized in Table~\ref{tab:Te127}. 
Fig.~\ref{fig:141Ndmg-Ce} shows the measured and calculated isomeric-cross section ratios as a function of the incident $^3He$-particle energy for the best fit with the $\eta =0.20$ and two other values ($\eta =0.05$, $\eta =0.95$) of $^{141}$Nd and the optimal value of $^{142}$Nd were fixed. 
 
The $^{139}$Nd and $^{141}$Nd isotopes can be produced by proton induced reaction on $^{141}$Pr too.
Fig.~\ref{fig:141Nd-Pr} depicts the results for the reduced $\chi ^2$ for comparing the model-calculated data and the experimental data for the $^{139}$Nd and $^{141}$Nd nuclei as a function of $\eta $ values of the $^{139}$Nd, $^{140}$Nd and $^{141}$Nd and number of the discrete levels of $^{139}$Nd and $^{141}$Nd. 
The reduced $\chi ^2$ as a function of the N levels is very similar for the $^3He$ and $p$ induced reactions in the case of the $^{141}$Nd in spite of that N=11 was used for the $^3He$ and N=4 for $p$ induced reaction. There is only a slight difference between the $\chi ^2$ values of the N=4 and N=11 in the case of the $^3He$ induced reaction.
The best estimation of $\eta$ values for the $^{139}$Nd, $^{140}$Nd and $^{141}$Nd nuclei and their comparison with the $\eta$ values derived from the discrete levels are summarized in Table~\ref{tab:Te127}. 

Figs.~\ref{fig:141Ndmg-Pr}a,b
 show the measured and calculated isomeric-cross section ratios as a function of the incident proton energy for $^{141}$Pr$(p,xn)^{139}$Nd$^{m,g}$ and $^{141}$Pr$(p,xn)^{141}$Nd$^{m,g}$ reactions, respectively, with
 the best fit $\eta $ and with two other values.

\subsection{$^{159}$Tb$(\alpha ,n)^{162}$Ho$^{m,g}$}

\begin{figure}
\centering\includegraphics*[scale=\ones ]{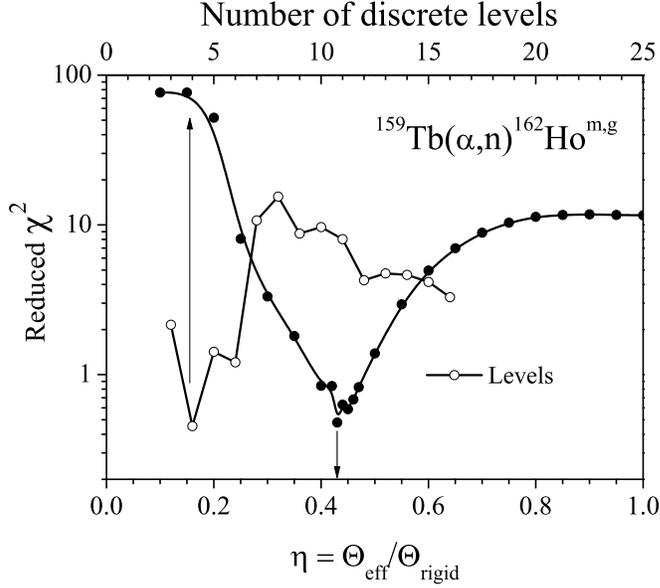}
\caption{\label{fig:162Ho} Reduced $\chi ^2$ as a function of the number of discrete levels and $\eta $ for the $^{159}$Tb$(\alpha ,n)^{162}$Ho$^{m,g}$ reaction. 
}
\end{figure}

\begin{figure}
\centering\includegraphics*[scale=\ones ]{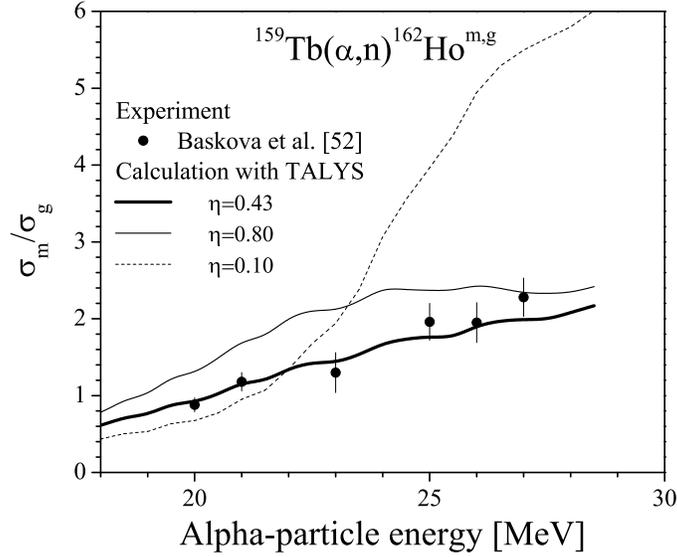}
\caption{\label{fig:162Homg} Measured and calculated isomeric cross-section ratios for the $^{159}$Tb$(\alpha ,n)^{162}$Ho$^{m,g}$ reaction using the optimal number of discrete levels and the $\eta $ values of 0.43, 0.10 and 0.80. 
}
\end{figure}

The experimental isomeric cross-section ratios were taken from \citet{Baskova1985}. 
Fig.~\ref{fig:162Ho} depicts the reduced $\chi ^2$ for comparing the model-calculated data with the experimental data.
This calculation was performed with the TALYS 1.6 code without using the default precomplex emission mode, and spin cut off factor energy dependence was treated as the original BSFG. 
The best fit to the experimental data as a function of the number of levels (N) was obtained at N=4. The decay scheme of the $^{162}$Ho contains only 16 discrete levels. 
The results of the reduced $\chi ^2$ as a function of the $\eta $ value can also be seen in Fig.~\ref{fig:162Ho}. The best fit of the experimental ratio can be achieved by $\eta =0.43\pm 0.06$ value. 
The best estimation of $\eta$ value for the $^{162}$Ho nucleus and its comparison with the $\eta$ value derived from the discrete levels are summarized in Table~\ref{tab:Te127}. Fig.~\ref{fig:162Homg} shows the measured and calculated isomeric cross-section ratios using the optimal number of discrete levels and the $\eta $ values of 0.10 and 0.80.

\subsection{$^{181}$Ta$(\alpha ,3n)^{182}$Re$^{m,g}$}

\begin{figure}
\centering\includegraphics*[scale=\ones ]{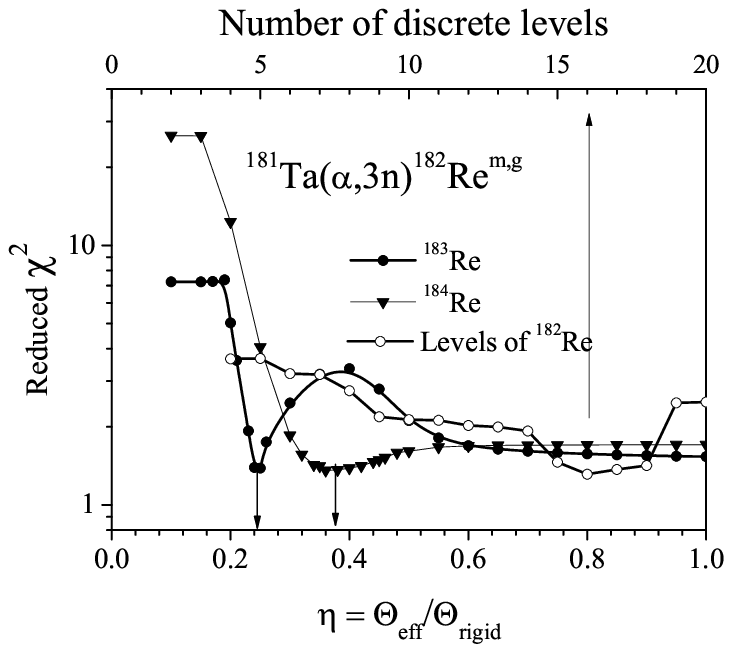}
\caption{\label{fig:182Re} Reduced $\chi ^2$ as a function of the number of discrete levels and $\eta $ for the $^{181}$Ta$(\alpha ,3n)^{182}$Re$^{m,g}$ reaction. 
}
\end{figure}

\begin{figure}
\centering\includegraphics*[scale=\ones ]{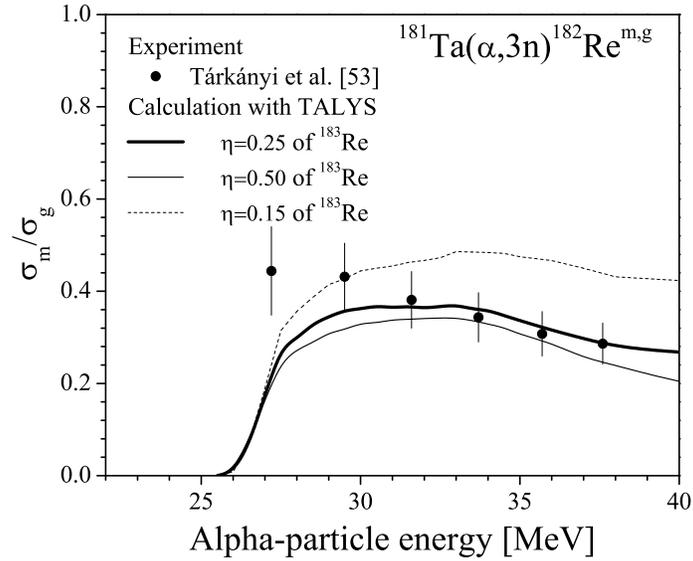}
\caption{\label{fig:182Remg} Measured and calculated isomeric cross-section ratios for the $^{181}$Ta$(\alpha ,3n)^{182}$Re$^{m,g}$ reaction using the optimal number of discrete levels and the $\eta $ values of 0.25, 0.15 and 0.50 for the $^{183}$Re. 
}
\end{figure}

The experimental isomeric cross-section ratios were calculated from the cross section data of \citet{Tarkanyi2003}. 
The formula \ref{eq:erreq} was used to calculate their uncertainties. Before the model calculation, the level scheme of the product $^{182}$Re was checked in the TALYS input files, and it was compared with ENSDF \cite{ensdf} library. 
Unfortunately, there is no gamma transition from the isomeric level to the ground state; therefore, the energy of the isomeric level is unknown. 
The TALYS level file set an energy of 0.1 MeV for the isomeric state, and the excited levels connected to the isomeric state were not included. 
Relatively good fit can be achieved setting the energy of the isomeric state to 0.550 MeV, and assuming that excited states decay to the isomeric state was involved.

 Both the optimal number of the used discrete levels and $\eta$ were determined by the procedure described above. Fig.~\ref{fig:182Re} shows the reduced $\chi ^2$ of the fitting of the model-calculated data to the experimental data as a function of the used discrete levels and $\eta$ values. 
 The minimum value of the ${\chi }^{2}$ corresponds to the optimal number of levels of 16. There does not exist a minimal value in the case of $^{182}$Re, probably due to the incorrect level scheme and the limited energy range of the experimental data. Minimal values were obtained for $^{183}$Re and $^{184}$Re nuclei. 
The best estimation of $\eta$ values for these nuclei and their comparison with the $\eta$ value derived from the discrete levels are summarized in Table~\ref{tab:Te127}.
 The isomeric cross-section ratio as a function of the $\eta $ was calculated and compared with the experimental data. Fig.~\ref{fig:182Remg} shows the measured and calculated isomeric cross-section ratios as a function of the incident $\alpha $-particle energy for three different $\eta$ values.

\subsection{$^{nat}$Pt$(^3He,xn)^{195}$Hg$^{m,g}$ and $^{nat}$Pt$(\alpha ,xn)^{197}$Hg$^{m,g}$ }

\begin{figure}
\centering\includegraphics*[scale=\ones ]{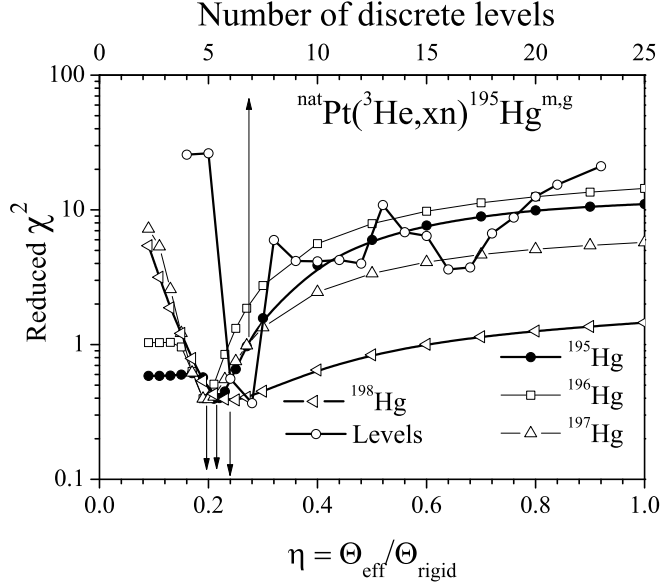}
\caption{\label{fig:195Hg} Reduced $\chi ^2$ as a function of the number of discrete levels and $\eta $ for $^{195}$Hg $^{196}$Hg $^{197}$Hg $^{198}$Hg from the $^{nat}$Pt$(^3He,xn)^{195}$Hg$^{m,g}$ reaction. 
}
\end{figure}

\begin{figure}
\centering\includegraphics*[scale=\ones ]{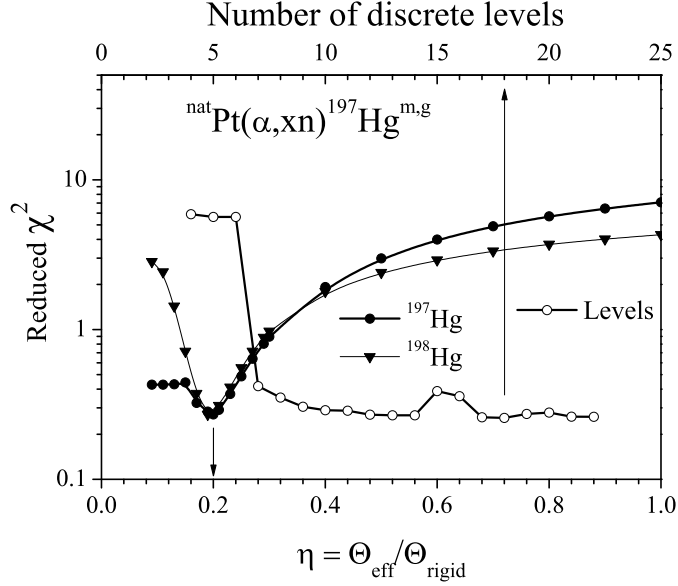}
\caption{\label{fig:197Hg} Reduced $\chi ^2$ as a function of the number of discrete levels and $\eta $ for the $^{nat}$Pt$(\alpha ,xn)^{197}$Hg$^{m,g}$ reaction.
}
\end{figure}

The cross sections, isomeric cross-section ratios and the analysis of the $\eta $ value were published in \cite{sudar2006}. Here we have reanalyzed the $\eta $ value using the framework presented in this paper. 
In the original publication, the STAPRE code was used for the model calculation and the best $\eta $ values were only estimated. Figures ~\ref{fig:195Hg} and ~\ref{fig:197Hg} show the results of the numerical analysis using the TALYS 1.6 code and the same level density parameters  as used in the STAPRE calculation. 
The best fit to the experimental data can be achieved by $\eta =0.19\pm 0.02$ for $^{195}$Hg and $\eta =0.20\pm 0.03$ for $^{197}$Hg. 
These data are in good agreement with the results presented in \cite{sudar2006}. This proves that the model calculation with the STAPRE or TALYS 1.6 does not have any important difference from this point of view.

\subsection{$^{197}$Au$(\alpha ,3n)^{198}$Tl$^{m,g}$}

\begin{figure}
\centering\includegraphics*[scale=\ones ]{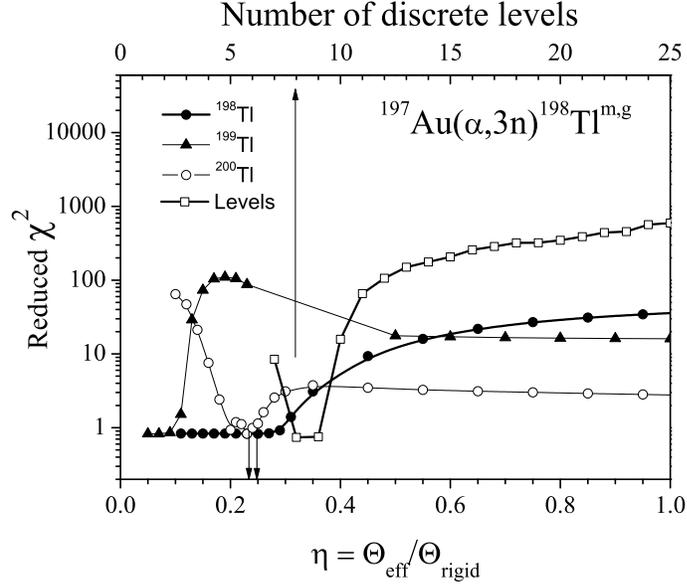}
\caption{\label{fig:198Tl} Reduced $\chi ^2$ as a function of the number of discrete levels and $\eta $ for $^{198}$Tl $^{199}$Tl and $^{200}$Tl nuclei in the $^{197}$Au$(\alpha ,3n)^{198}$Tl$^{m,g}$ reaction. 
}
\end{figure}

\begin{figure}
\centering\includegraphics*[scale=\ones ]{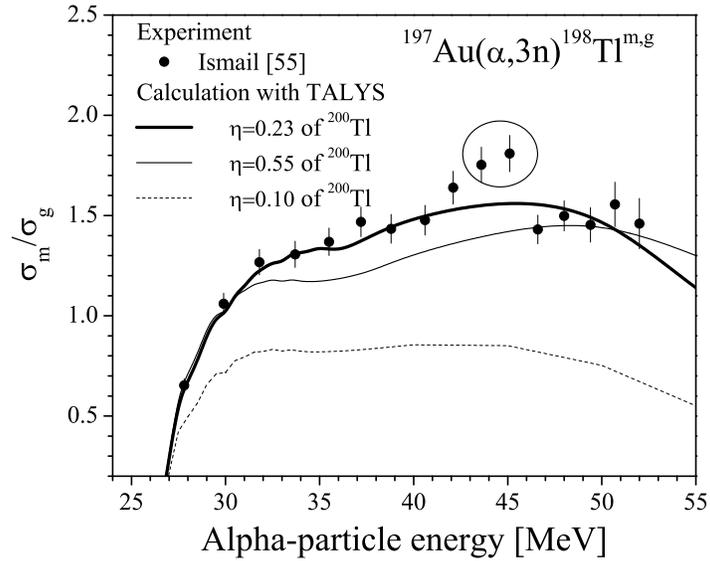}
\caption{\label{fig:198Tlmg} Measured and calculated isomeric cross-section ratios for the $^{197}$Au$(\alpha ,3n)^{198}$Tl$^{m,g}$ reaction using the optimal number of discrete levels and the $\eta $ values of 0.23, 0.10 and 0.50 of the $^{200}$Tl nucleus. The experimental points were removed from the fit are encircled.
}
\end{figure}

There are many measurements for the $^{197}$Au$(\alpha ,3n)^{198}$Tl$^{m,g}$ reaction, but the isomeric cross-section ratios are very contradictory. Therefore the latest data by \citet{Ismail1998} were selected, covering a wide energy range and with explicit numerical values. 
For those data the experimental isomeric cross-section ratios were calculated. Regarding the uncertainties, besides the statistical error, the overall uncertainties were also included. Unfortunately the level scheme of $^{198}$Tl is very uncertain. 
Most of the levels have only suggested spin and parity values. The level number 7 is the isomeric state with defined 7$^+$ spin and parity. 
The levels 10, 11 and 12 decay to the isomeric state completely according to the decay scheme. The spin and parity of these levels are (6)$^+$, (5$^+$) and (10$^-$), respectively (the brackets indicate the suggested values). 
Fig.~\ref{fig:198Tl} shows the reduced $\chi ^2$ as a function of the number of discrete levels. 
The reduced $\chi ^2$ increases to a very high value of more than 100 for the level 12 and it does not decrease any more. The usual change is about a factor of 10 for all the previously presented cases, which indicates that the spin and parity and branching ratios are unreliable for levels 11 and 12.
The effect of parameters of level 11 and 12 on the reduced $\chi ^2$ was tested. 
Finally N=8 was used in the calculations while determination of the level parameters could not be done unambiguously. 
Fig.~\ref{fig:198Tl} shows also the reduced $\chi ^2$ as a function of the parameter $\eta$ of $^{198}$Tl, $^{199}$Tl and $^{200}$Tl nuclei. 
The best fit to the experimental data is presented in Table~\ref{tab:Te127}. Unfortunately only an upper limit can be estimated for the $^{199}$Tl because  a defined minimal value was not found. In the calculation of the $\chi ^2$ some data points which were out of the trend of the data were excluded. Fig.~\ref{fig:198Tlmg} depicts the isomeric cross-section ratio as a function of the incident $\alpha $-particle energy at the best fit $\eta =0.23$ as well as at $\eta =0.10$ and $\eta =0.50$ for the $^{200}$Tl nucleus.

\section{Discussion}

\begin{figure}
\centering\includegraphics*[scale=\ones ]{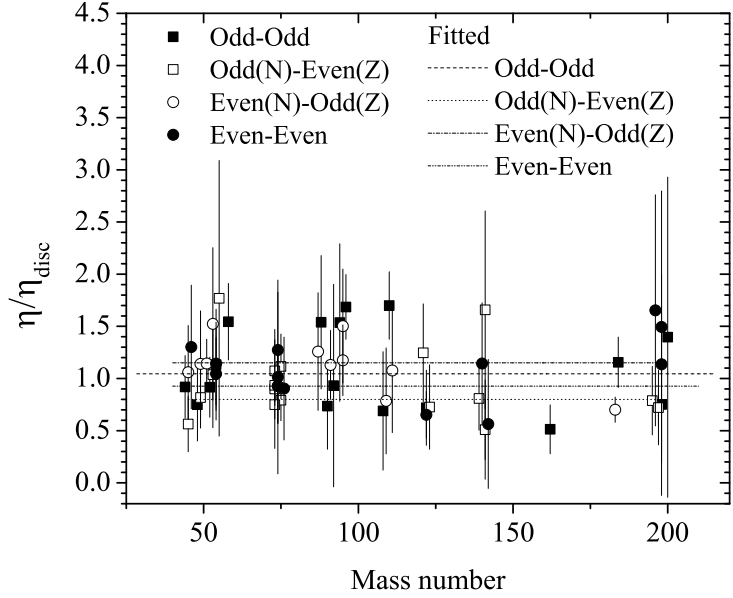}
\caption{\label{fig:etacd-A} $\eta $/$\eta _d$ as a function of the mass number of the investigated nuclei.
}
\end{figure}

\begin{figure}
\centering\includegraphics*[scale=\ones ]{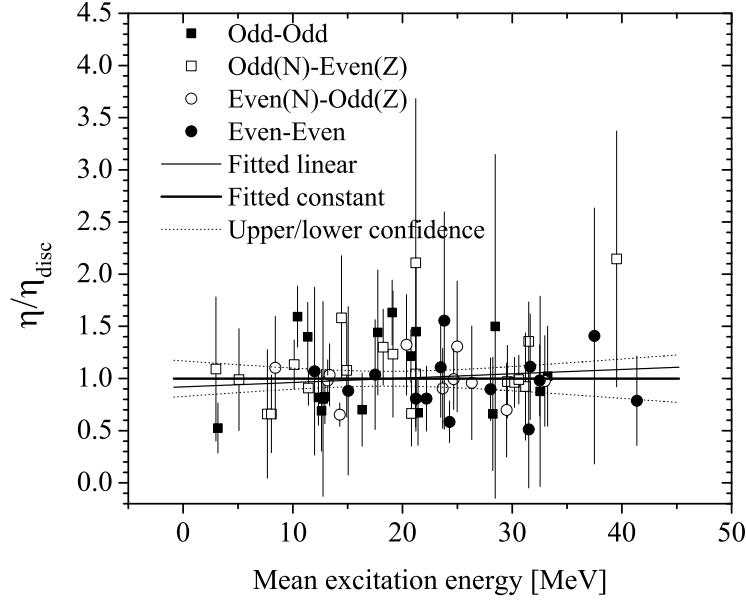}
\caption{\label{fig:etacd-E} $\eta $/$\eta _d$ as a function of the excitation energy of the investigated nuclei. The fitted linear function and the constant function with its upper/lower 95$\% $ confidence level are shown.
}
\end{figure}

\begin{table*}
\caption[]{\label{tab:Fit-etacdx}Parameters of the fitted functions 
}
\begin{ruledtabular}
\begin{tabular}{ccccccccccc}
Fitted function & & a+b*x & & & & &$\mid $& & c & \\
\hline
x=A \\
\hline
Type & A & $\Delta a$ & b & $\Delta b$ & Reduced $\chi ^2$ & Prob(b=0){[}\%{]} &$\mid $& c & $\Delta c$ & Reduced $\chi ^2$ \\
\hline
Odd-Odd & 1.18 & 0.24 & -0.0012 & 0.0020 & 1.28 & 55 &$\mid $& 1.05 & 0.10 & 1.23 \\
Odd(N)-Even(Z) & 0.88 & 0.09 & -0.0008 & 0.0008 & 0.41 & 33 &$\mid $ & 0.80 & 0.04 & 0.42 \\
Even(N)-Odd(Z) & 1.20 & 0.17 & -0.0007 & 0.0022 & 0.16 & 75 &$\mid $& 1.15 & 0.05 & 0.14 \\
Even-Even & 1.13 & 0.17 & -0.0022 & 0.0017 & 0.29 & 22&$\mid $ & 0.93 & 0.07 & 0.30 \\
\hline
x=E$_{exc}$ &\\
\hline
All data & 0.92 & 0.09 & 0.0041 & 0.0041 & 0.55 & 31 &$\mid $& 1.0 & 0.04 & 0.54\\
\end{tabular}
\end{ruledtabular}
\end{table*}

\subsubsection{Excitation energy and mass dependence of the $\eta $/$\eta _d$ values} 

The Tables~\ref{tab:Sc-V-cd}-\ref{tab:Te127} show that the evaluated $\eta $ values
cover the  range of 0.15-1.13 while the $\eta $/$\eta _d$ data are scattered around 1. 
Fig.~\ref{fig:etacd-A} depicts the $\eta $/$\eta _d$ values as a function of the mass number A. 
It does not show any visible mass dependence, but to test the possible mass dependence, a simple linear function was fitted
 to the data separately for the odd-odd, odd N - even Z, even N - odd Z and even-even type of nuclei. The fitting procedure was the least squares method, with weighting  by inverse square of the uncertainties of the data.
 After the first fit those data which were farther from the average, represented by the fitted curve, by more than three times their uncertainty, were removed from the evaluation set (3 data points), and the fitting procedure was repeated. 
The results are given in  Fig.~\ref{fig:etacd-A}. The $\eta $/$\eta _d$ ratio seems to be constant as a function of the mass number, but slightly depends on the type of nuclear structure. 
Table~\ref{tab:Fit-etacdx} presents the fitted parameter values using linear mass 
dependence function and assuming a constant for the mass dependence. 
It can be seen that the uncertainty of the slope is higher than the value of the slope in three cases from the four. 
This indicates that the slope is probably zero. 
We have done a further test to prove our assumption. 
The Student's t-test was calculated for the data and the probability of the slope equal to zero at 95$\% $ confidence level is presented in the Table~\ref{tab:Fit-etacdx}. 
Those data also show that the slope is zero with large probability. It is hard to prove from experimental data that a value is zero because of the experimental uncertainties. 
To be sure that the slope is not zero the absolute value of the slope should be 3-5 times higher than its uncertainties. 
Therefore we can state that there is no proven evidence that the slopes are not zero. 
Finally, we propose to accept the slope zero and the fitted constant values. 
This means that the $\eta_d$ for the discrete levels follows the same mass dependence as $\eta $ value for the continuum; therefore it is the inherent property of the nuclear structure.

This result means that the $\eta $ value can be estimated from the analysis of the discrete levels for any nucleus multiplying the analyzed $\eta _d$ value with the fitted parameter of the appropriate type of the nuclei.
{
The reduced $\chi ^2$ of the fitted constants is near to one or less than one, and the uncertainty of the constants is less than 10 percent. The effect of the untreated nuclear reaction model parameters should appear as mass number dependence, because most of them depend on the mass number or increase the scattering of the evaluated $\eta $ value.  Therefore, the effect of the untreated nuclear model parameters can be estimated to be less than 10 percent. 
}

The Tables~\ref{tab:Sc-V-cd}-\ref{tab:Te127} contain two columns presenting the excitation energy of the nucleus which may contribute to the investigated isomeric cross-section ratio. 
As far as we know, this is the first case where  $\eta $ values were evaluated above the neutron separation energies. Introducing the average excitation energy as the mean of minimal and maximal excitation energy, the energy dependence of the $\eta $/$\eta _d$ ratio can be analyzed. 
 The $\eta $/$\eta _d$ ratios were normalized with the constant fit values determined above to remove the effect of the isotope type. 
 We do not have enough information to analyze the energy dependence for the individual nuclei, but Fig.~\ref{fig:etacd-E} reveals the very interesting behavior of the $\eta $/$\eta _d$ values which seem to be constant as a function of the mean excitation energy. 
 A linear and a constant type functions were fitted to each data set as described above. Table~\ref{tab:Fit-etacdx} presents the fitted parameter values for the energy dependence too. For similar reason as above, the slope of the linear function can be set to zero and the constant fit can be accepted. 
 Fig.~\ref{fig:etacd-E} shows the fitted linear function and the constant function with its upper/lower 95$\% $ confidence level. 
 While constant fit gives the same quality description as the linear one, it seems appropriate to accept the simplest function for the description of the data. 
 This means that the $\eta $/$\eta _d$ in averages is  independent of the mean excitation energy in the continuum. 
 Since $\eta _d$ is independent of the excitation energy in the continuum by its definition, therefore it follows that  $\eta $  is also independent of the excitation energy. 
 
The energy dependence of the $\eta _d$ values was investigated on 15 nuclei. There were only a few cases when the $\eta _d$ values showed a defined trend instead of the constant values. 
Taking into account the significant uncertainties of the individual points, those trends can be overridden with a constant value too. In the case of $^{54}$Fe, there were three points up to 20 MeV, which proved the constant value of $\eta $. 
Therefore we state with a large probability that the $\eta$ values seem to be independent of the excitation energy.
The introduction of the average excitation energy as the mean of minimal and maximal excitation energy is arbitrary, but in the light of the final result, it is unimportant. 
The interpolation of the spin-cutoff factor in TALYS  introduced an indirect energy dependence for $\eta $ which has not disturbed the situation, because most of the data were above the neutron separation energy.

\subsubsection{Mass and (N-Z) dependence of $\eta $}
 
\begin{table*}
\caption[]{\label{tab:Fit-eta-odd-a}Parameters of the fitted functions for odd mass nuclei 
}
\begin{ruledtabular}
\begin{tabular}{ccccccccccc}
Odd mass & & $45\le A\le 199$ & & & \\
\hline
Fitted function & & $a*e^{ (-b*A)}$ & & & \\
\hline
Type & a & $\Delta a$ & b & $\Delta b$ & Reduced $\chi ^2$ \\
\hline
Odd(N)-Even(Z) & 1.42 & 0.20 & 0.0108 & 0.0017 & 3.32 \\
Even(N)-Odd(Z) & 1.45 & 0.23 & 0.0091 & 0.0018 & 3.72 \\
All odd mass data & 1.41 & 0.15 & 0.0099 & 0.0012 & 3.56 \\
\hline
Fitted function & & $a*e^{ (-b*(N-Z))}$ & & & \\
\hline
Odd(N)-Even(Z) & 0.84 & 0.06 & 0.0443 & 0.0068 & 3.05 \\
Even(N)-Odd(Z) & 1.08 & 0.08 & 0.0475 & 0.0055 & 2.07 \\
All odd mass data & 0.90 & 0.05 & 0.0430 & 0.0045 & 3.08
\end{tabular}
\end{ruledtabular}
\end{table*}
 
\begin{table*}
\caption[]{\label{tab:Fit-eta-even-a}Parameters of the fitted functions for even mass nuclei 
}
\begin{ruledtabular}
\begin{tabular}{cccccccccccc}
Odd-odd & & & & & & \\
range & Fitted function & a & $\Delta a$ & b & $\Delta b$ & Reduced $\chi ^2$ \\
\hline
$44\le A\le 110$ & a & 0.97 & 0.03 & & & 1.12 \\
$110< A\le 198$ & a*exp(-b*A) & 5.81 & 0.87 & 0.0160 & 0.0008 & 0.35 \\
$(N-Z)\le 13$ & a & 1.07 & 0.03 & & & 0.43 \\
$13<(N-Z)$ & a*exp(-b*(N-Z)) & 2.32 & 0.44 & 0.0581 & 0.0055 & 1.17 \\
\hline
Even-even & & & & & & \\
range & Fitted function & a & $\Delta a$ & b & $\Delta b$ & Reduced $\chi ^2$ \\
\hline
$44\le A\le 142$ & a*exp(-b*A) & 1.36 & 0.21 & 0.0138 & 0.0017 & 2.96 \\
$142< A\le 156$ & a+b*A & -0.93 & 0.54 & 0.0088 & 0.0036 & 0.90 \\
$156< A\le 198$ & a*exp(-b*A) & 6.98 & 2.93 & 0.0175 & 0.0023 & 0.42 \\
$(N-Z)\le 20$ & a*exp(-b*(N-Z)) & 0.72 & 0.09 & 0.0562 & 0.0090 & 4.54 \\
$20< (N-Z)\le 28$ & a+b*(N-Z) & -0.223 & 0.065 & 0.0230 & 0.0027 & 1.41 \\
$28< (N-Z)\le 38$ & a*exp(-b*(N-Z)) & 2.60 & 1.34 & 0.0646 & 0.0147 & 0.91
\end{tabular}
\end{ruledtabular}
\end{table*}

\begin{figure}
\centering\includegraphics*[scale=\ones]{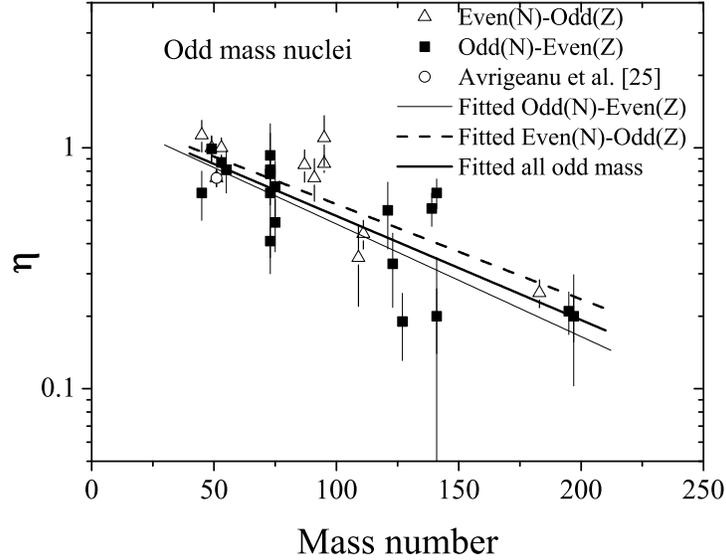}
\caption{\label{fig:etaoddA} Evaluated $\eta $ as a function of mass number, showing trends for odd mass number of nuclei. An exponentially decreasing trend was found for the odd mass  nuclei. 
}
\end{figure}

\begin{figure}
\centering\includegraphics*[scale=\ones]{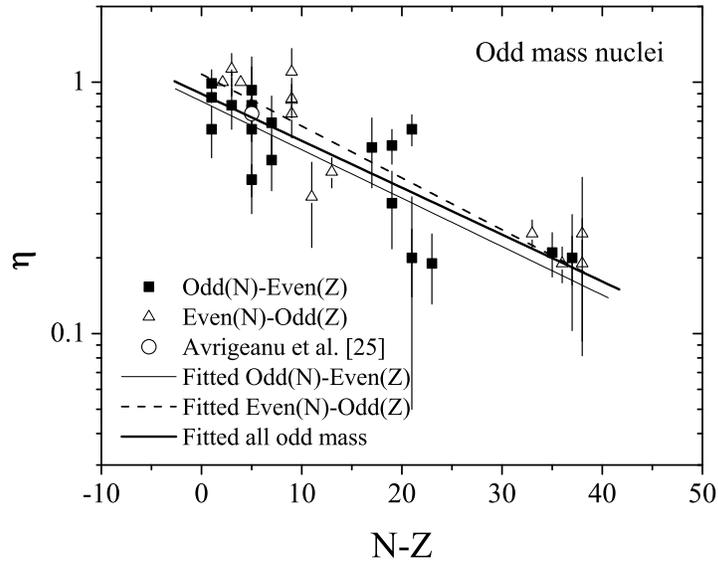}
\caption{\label{fig:etaoddNZ} Evaluated $\eta $ as a function of N-Z value of the nuclei, showing similar, exponentially decreasing trend for odd mass nuclei. 
}
\end{figure}

\begin{figure}
\centering\includegraphics*[scale=\ones]{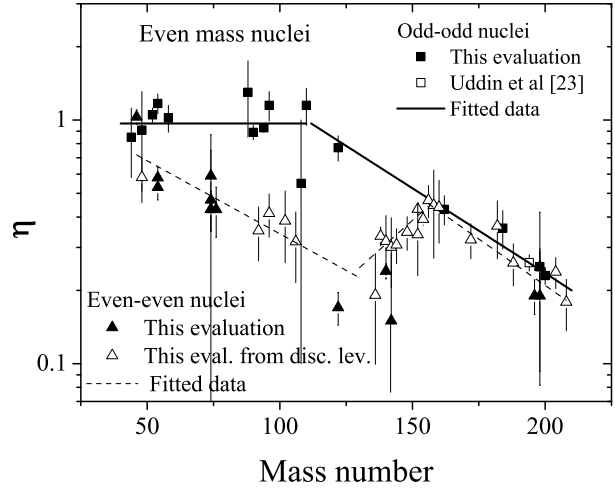}
\caption{\label{fig:etaevenA} Evaluated $\eta $ as a function of mass number, showing trends for even mass number of nuclei. The data derived from evaluation of discrete levels are shown by separate symbols. 
}
\end{figure}

\begin{figure}
\centering\includegraphics*[scale=\ones]{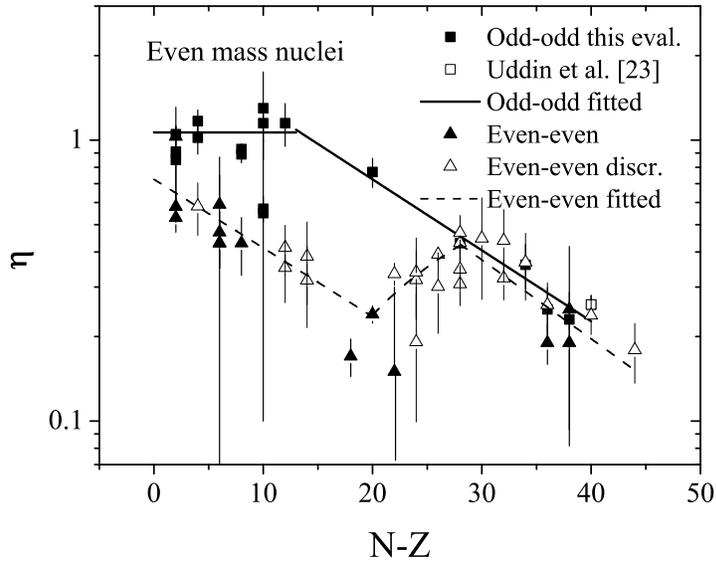}
\caption{\label{fig:etaevenNZ} Evaluated $\eta $ values as a function of N-Z value of the nuclei.
}
\end{figure}

\begin{table}
\caption[]{\label{tab:Disc-eta-even-even}Even-even $\eta $ values from discrete levels
} 
\begin{ruledtabular}
\begin{tabular}{ccccccccccc}
Nucleus & A & Z & N-Z & $\eta _d$ & $\Delta \eta _d$ & \\
\hline
$^{48}Ti$ & 48 & 22 & 4 & 0.58 & 0.12 & \\
$^{92}Zr$ & 92 & 40 & 12 & 0.35 & 0.09 & \\
$^{96}Mo$ & 96 & 42 & 12 & 0.41 & 0.08 & \\
$^{102}Ru$ & 102 & 44 & 14 & 0.39 & 0.13 & \\
$^{106}Pd$ & 106 & 46 & 14 & 0.32 & 0.10 & \\
$^{136}Ba$ & 136 & 56 & 24 & 0.19 & 0.09 & \\
$^{138}Ce$ & 138 & 58 & 22 & 0.33 & 0.03 & \\
$^{140}Ce$ & 140 & 58 & 24 & 0.32 & 0.09 & \\
$^{142}Ce$ & 142 & 58 & 26 & 0.30 & 0.10 & \\
$^{144}Ce$ & 144 & 58 & 28 & 0.31 & 0.05 & \\
$^{148}Nd$ & 148 & 60 & 28 & 0.35 & 0.05 & \\
$^{152}Sm$ & 152 & 62 & 28 & 0.43 & 0.02 & \\
$^{152}Gd$ & 152 & 64 & 24 & 0.34 & 0.11 & \\
$^{154}Gd$ & 154 & 64 & 26 & 0.39 & 0.02 & \\
$^{156}Gd$ & 156 & 64 & 28 & 0.47 & 0.07 & \\
$^{158}Gd$ & 158 & 64 & 30 & 0.45 & 0.18 & \\
$^{160}Gd$ & 160 & 64 & 32 & 0.44 & 0.13 & \\
$^{172}Yb$ & 172 & 70 & 32 & 0.32 & 0.05 & \\
$^{182}W $ & 182 & 74 & 34 & 0.37 & 0.10 & \\
$^{188}Os$	& 188 & 76 & 36 & 0.26	& 0.05 \\
$^{204}Pb$	& 204 &	82 & 40 & 0.24 & 0.03 \\
$^{208}Pb$	& 208 &	82 & 44 & 0.18 & 0.04 \\
\end{tabular}
\end{ruledtabular}
\end{table}

The evaluated data for nuclei in various mass regions  summarized in Tables~\ref{tab:Sc-V-cd}-\ref{tab:Te127}. %
  were later grouped according to the even/odd properties of the  mass number and neutron (N)and proton (Z) number of the nucleus. Fig.~\ref{fig:etaoddA} depicts the $\eta $ values of nuclei with an odd mass number. 
The odd N, even Z and even N, odd Z forms of the odd nuclei show quite similar behavior. In both cases the $\eta $ values of the nuclei show a linear trend on the semi-log plot in Fig.~\ref{fig:etaoddA}; therefore an exponential function, $a*e^{(-bA)}$, seemed appropriate to fit the data. 
The parameters of the fitted function with their uncertainties and the reduced $\chi ^2$ of the fit are presented in Table~\ref{tab:Fit-eta-odd-a}.
While data of the two types of odd mass number overlapped each other, the fitted functions are quite near and parallel; in fact it seems appropriate to fit a joint function for all odd data. (Level density systematics treat these two types in the same way too.) This fitted function and its parameters are also presented in Fig.~\ref{fig:etaoddA} and Table~\ref{tab:Fit-eta-odd-a}.

\citet{Nomura2011} have shown that the moment of inertia of the even-even isotopes of tungsten decreases with the increasing neutron number, both for experimental systematics and nuclear models. 
Therefore we tested also the behavior of $\eta $ values as a function of the (N-Z). The $\eta $ values of odd mass isotopes are plotted in Fig.~\ref{fig:etaoddNZ} whereby an exponentially decreasing trend is observed. 
The parameters of the fitted function with their uncertainties and the reduced $\chi ^2$ of the fit are also given in Table~\ref{tab:Fit-eta-odd-a}. 
The reduced $\chi ^2$ values are somewhat lower for the (N-Z) dependence fits as compared to the A dependence. 
If we analyze the deviation of the evaluated $\eta $ values and the fitted functions, then only 5 data give significant deviation from the fitted functions. 
If those data would be removed then the reduced $\chi ^2$ values would be near to one. 
However, presently there is no physical model which could describe the dependence of the $\eta $ values. Therefore, we do not have any objective reason to remove those data. 

Both A and N-Z dependence describe well the evaluated $\eta $ values, which are near to the stability line in the N, Z plane. Selecting a fixed Z number and increasing or decreasing the N number by two, the four calculated values will be different while the fitted exponential factor for the (N-Z) is about four times of the A dependence. Maybe this deviation helps to distinct which formula is more appropriate.

The $\eta $ values for the nuclei with  even mass number are presented in Fig.~\ref{fig:etaevenA} as a function of the mass number, both for the odd-odd and even-even types.
Simple exponential curves do not describe them as in the case of odd mass data. The $\eta $ values of the odd-odd nuclei below the mass number  110 seem to be constant, while those above that figure appear to follow a similar trend as the data for the odd mass number. 
The fitted function and its parameters are also presented in Fig.~\ref{fig:etaevenA} and Table~\ref{tab:Fit-eta-even-a}. 

The fit using these two functions is acceptable because the reduced $\chi ^2$ values are 1.12 and 0.35, respectively. The evaluated $\eta $ value of 0.97$\pm$0.03 is quite near to the value of 1. 
It seems to prove that use of the rigid body moment of inertia has been a good estimate for odd-odd type nuclei below mass number 110. The characteristic constant above mass number 110 is higher than the same value for the nuclei with an odd mass number.

The behavior of the even-even type nuclei is different from the odd-odd type ones. It shows an exponentially decreasing trend starting even from the lowest value. 
Unfortunately, there are relatively few evaluated data for this type of nuclei. Therefore it was considered a good opportunity to extend the data with the $\eta $ values derived from the discrete levels. 
They are plotted in Fig.~\ref{fig:etaevenA} too and the numerical data are given in Table~\ref{tab:Disc-eta-even-even}. 
The new data fit well to the evaluated data up to mass number 142. There is a large gap in the $\eta $ value evaluated from the isomeric cross-section ratio for nuclei with mass numbers between 142 and 196. 
The new data for this range show an unexpected behavior: 
The $\eta $ values increase and reach the systematics of the odd-odd type data at the $^{156}Gd$, then they follow the trend of the odd-odd type of data. 
We interpret this result by assuming closed shell structures:  magic number of 126 nuclei (63 proton-neutron pairs), 28 neutrons build up a closed shell structure, and the additional one neutron and a proton get into the same role as they have in the odd-odd type structure. The parameters of the fitted functions are presented in Table~\ref{tab:Fit-eta-even-a}. It is seen that the characteristic constants of the three exponentias are in good agreement within their error limits but differ significantly  from the data of odd mass nuclei.

The $\eta $ values as a function of the (N-Z) for even mass isotopes are plotted in Fig.~\ref{fig:etaevenNZ}. This figure represents a similar behavior as the mass number dependent plot. The $\eta $ values seem to be constant up to (N-Z)=13 for the odd-odd type of nuclei. The fitted constant $1.07\pm 0.3$  agrees quite well with the fitted value of the mass dependence. Above (N-Z) of 13 it shows an exponential decrease, but a higher characteristic constant as compared to the mass dependence. The $\eta $ values for the even-even type of nuclei show a similar trend as  the mass dependence, namely an exponential decrease up to (N-Z) equal to 20, a linear increase up to 28 and then following the trend of the odd-odd type nuclei. 

\subsubsection{Level density model dependence}

\begin{table*}
\caption[]{\label{tab:lvmodel}Test of the effect of the level density model on the evaluation of   $\eta $ using the $^{111}$Cd(p,2n)$^{110}$In and $^{141}$Pr(p,3n)$^{139}$Nd reactions. Constant Temperature Model (CTM) and the Generalized Superfluid Model (GSM) are compared with the Back-shifted Fermi-Gas Model (BFM).
}
\begin{ruledtabular} 
\begin{tabular}{cccccccccccccc}
Isotop & $^{110}$In &  & $^{111}$In &  &  & $^{139}$Nd &  & $^{140}$Nd &  & $^{141}$Nd&  &  \\
 &$\eta $&$\Delta \eta $&$\eta $&$\Delta \eta $&Red.$\chi ^2$& $\eta $ & $\Delta \eta $&$\eta $&$\Delta \eta $&$\eta $& $\Delta \eta $&Red.$\chi ^2$\\
\hline
CTM & 1.18 & 0.22 & 0.46 & 0.03 & 1.43 & 0.65 & 0.09 & 0.24 & 0.02 & 0.63 & 0.09 & 1.52 \\
BFM & 1.15 & 0.2 & 0.44 & 0.02 & 1.3 & 0.56 & 0.09 & 0.24 & 0.02 & 0.65 & 0.09 & 1.54 \\
GSM & 1.16 & 0.29 & 0.44 & 0.05 & 1.23 & 0.61 & 0.08 & 0.24 & 0.02 & 0.63 & 0.08 & 1.51 \\
\hline
 &$\frac{\eta }{\eta }$&$\Delta (\frac{\eta }{\eta })$&$\frac{\eta }{\eta }$&$\Delta (\frac{\eta }{\eta })$&  &$\frac{\eta }{\eta }$&$\Delta (\frac{\eta }{\eta })$& $\frac{\eta }{\eta }$&$\Delta (\frac{\eta }{\eta })$&$\frac{\eta }{\eta }$&$\Delta (\frac{\eta }{\eta })$& \\
CTM/BFM & 1.03 & 0.26 & 1.05 & 0.08 &  & 1.16 & 0.25 & 1.00 & 0.12 & 0.97 & 0.19 &  \\
GSM/BFM & 1.01 & 0.31 & 1.00 & 0.12 &  & 1.09 & 0.23 & 1.00 & 0.12 & 0.97 & 0.18 & \\
\hline
EMPIRE&&&&&Red.$\chi ^2$&&&&&&&Red.$\chi ^2$\\
\hline
EGSM&&&&& 20.9(2.3) &&&&&&& 10.4\\
GSM&&&&& 34.6(1.2) &&&&&&& 10.2\\
\hline
\end{tabular}
\end{ruledtabular} 
\end{table*}

\begin{figure}
\centering\includegraphics*[scale=1.0]{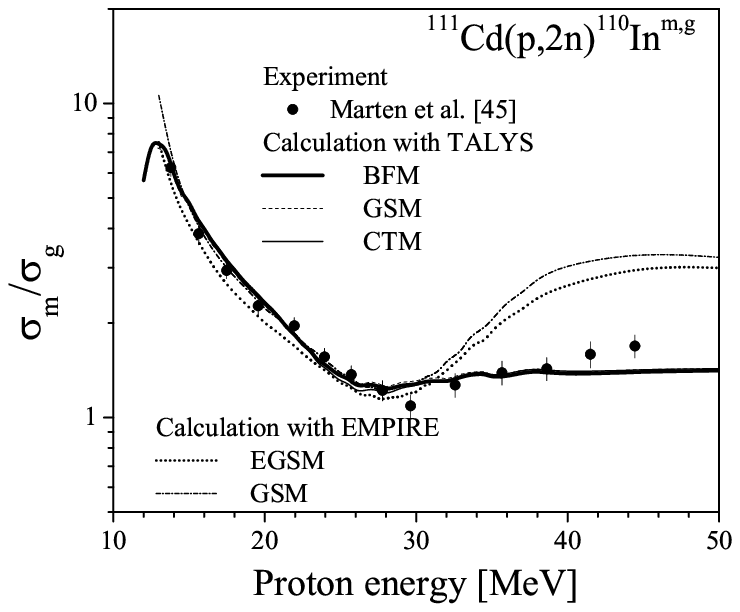} 
\caption{\label{fig:empIn110} Comparing the measured and calculated isomeric cross-section ratios for $^{111}$Cd(p,2n)$^{110}$In$^{m,g}$ reaction using different level density models of the TALYS 1.6 and the EMPIRE 3.27. The results of the three level density models of TALYS are overlapped with each other in this scale.  
}
\end{figure}
\begin{figure}
\centering\includegraphics*[scale=1.0]{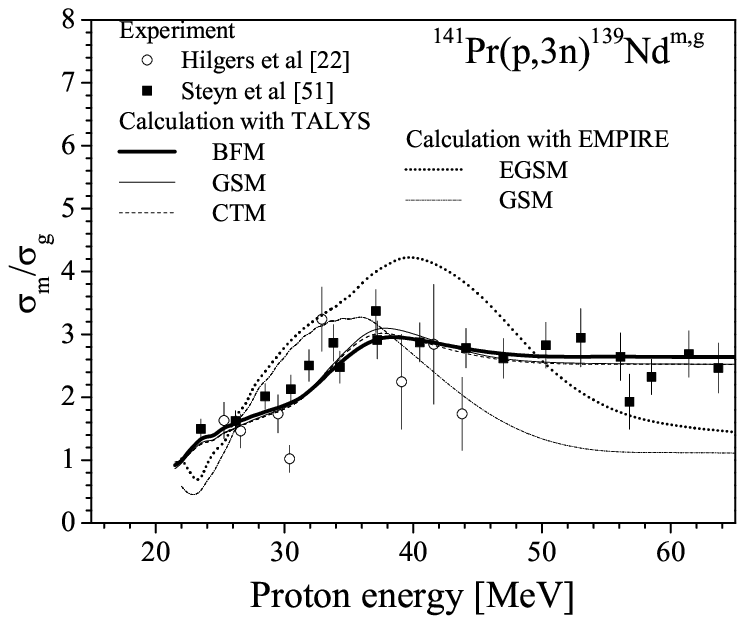}%
\caption{\label{fig:empNd139} Comparing the measured and calculated isomeric cross-section ratios for $^{141}$Pr(p,3n)$^{139}$Nd$^{m,g}$ reaction using different level density models of the TALYS 1.6 and the EMPIRE 3.27. The results of the three level density models of TALYS are overlapped with each other in this scale. 
}
\end{figure}

In the evaluation process above, the back-shifted Fermi gas model was used for calculation of the level density. In order to investigate how these results depend on  the level density model, two reactions were selected to compare the effect of the model, namely $^{111}$Cd(p,2n)$^{110}$In and $^{141}$Pr(p,3n)$^{139}$Nd.  The result of the calculations is summarized in Table \ref{tab:lvmodel}. All the three types of level density models use the same function for the spin dependence, Eq. (\ref{eq:spindep}). The evaluated $\eta $ values  agree quite well, indicating that the established systematics is independent of the  level density model used. The reduced $\chi ^2$ values are quite similar; therefore it can not be said that one or the other  model describes  the experimental data better from this point of view.  
An important result of this evaluation is that the $\eta $ value is independent of the excitation energy. There are approaches where $\eta $ or effective momentum of inertia is treated as an energy-dependent quantity. For example, the EMPIRE code \cite{Herman}  uses the Generalized Superfluid Model (GSM) with the original approach of Ignatyuk \cite{ignatyuk1975}, and the EGSM  model  uses, as GSM, the super-fluid model below the critical excitation energy and the Fermi Gas model above. Ground state deformation is damped with the increasing
nuclear temperature since it is known that  nuclei  become spherical at high excitation energies.
Moments of inertia for the yrast states  are calculated for deformation
using expressions proposed by Vigdor and Karwowski \cite{vigdor1982}. Both the GSM  and EGSM model implicate an energy dependence of  $\eta $  and in both cases at high energy they tend to near to one (depending on the deformation of the nucleus).  

Therefore, it is an important question how these models are able to describe the isomeric cross-section ratios. We did not want to do a full test of these models, therefore, the calculation was carried out only for those two cases which were used in this comparison above. 

The result is summarised in Table \ref{tab:lvmodel} too. To make  the result more clear,  comparisions of the measured and calculated  isomeric cross-section ratios are also presented.  Fig.~\ref{fig:empIn110} shows isomeric cross-section ratios for the $^{111}$Cd(p,2n)$^{110}$In$^{m,g}$  reaction. In this case the $\eta $ value for the $^{110}$In is $1.15\pm 0.2$ which is near to the 1.0, i.e. $\Theta _{eff}$ equal to $\Theta _{rig}$.
 It is visible that TALYS describes the isomeric cross-section ratio over the full energy range, but the EMPIRE is able to describe it neither with  the EGSM nor with the GSM model over the full energy range. 
Their result is quite good up to 29.63 MeV but above it, the calculated data are much higher. One explanation could be that the $\eta $ value for $^{111}$In is only $0.44\pm 0.02$. But the higher  $\eta $ value for $^{111}$In cross section  would decrease the isomeric cross-section ratios according to the TALYS calculation. 
Table \ref{tab:lvmodel} presents the reduced $\chi ^2$ for the full energy range and in the bracket up to 29.63 MeV. It can be seen that the GSM gives a better value for the reduced $\chi ^2$ than the EGSM model, and the GSM value up to 29.63 MeV agrees with the values for the TALYS evaluation.  But the  reduced $\chi ^2$ are very high in the full energy range and indicate that the model is not able to describe the isomeric cross-section ratios.

Fig.~\ref{fig:empNd139} presents the isomeric cross-section ratios for the  $^{141}$Pr(p,3n)$^{139}$Nd$^{m,g}$ reaction. 
In this case,  $\eta $ values significantly less than 1 are needed to describe the isomeric cross-section ratios. It is visible that the EMPIRE is not able to describe the isomeric cross-section ratios over any energy region. We think it would be proper to revise the concept of the spin distribution in the EMPIRE after a more wide comparison to the experimental isomeric cross-section ratios has been done.

\subsubsection{Influence on the total production cross section}

\begin{figure}%
\includegraphics*[scale=\ones]{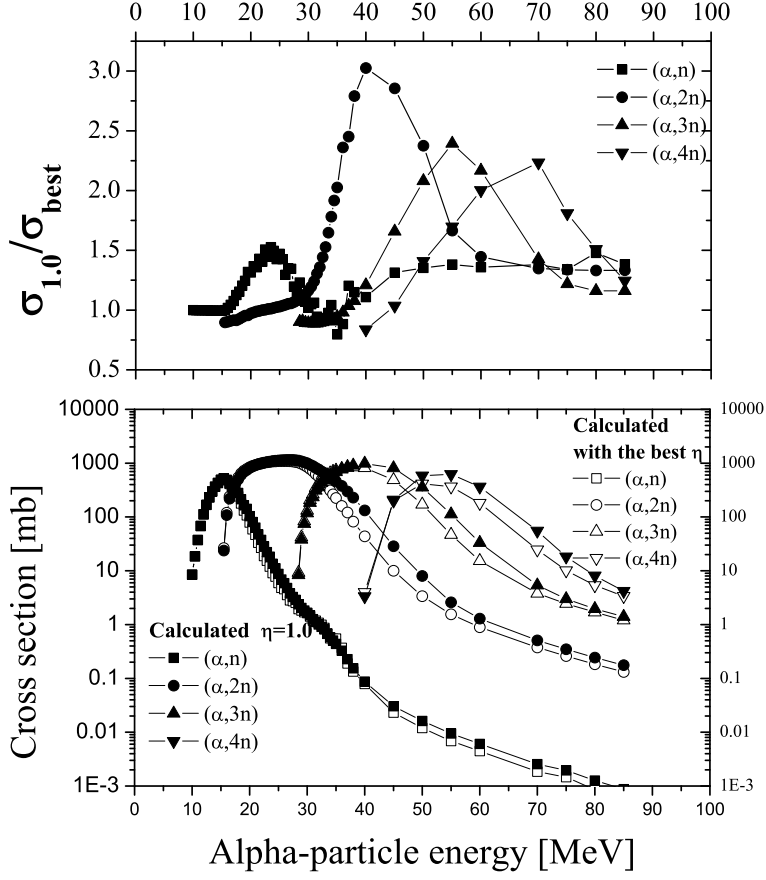}
\caption{\label{fig:eta-effec-cs} The  effect of $\eta $ on the calculated activation cross section taking into account  the differences of odd-even mass characteristics of $\eta $. The lower part of the figure shows the calculated cross sections for the $^{89}$Y($\alpha$,xn) reactions with mass independent $\eta =1.0$ and the best $\eta $ value obtained from the isomeric cross-section ratio for the $^{89}$Y$(\alpha ,3n)^{90}$Nb$^{m,g}$ reaction. The upper part of the diagram gives the ratio of the cross sections calculated using the two $\eta $ values.
}
\end{figure}

To check the influence of the nuclear mass effect in $\eta $ on the model calculation, activation cross sections were calculated for the $^{89}$Y($\alpha$,xn)$^{92,91,90,89}$Nb reactions using on one hand constant $\eta =1.0$ for all isotopes and, on the other, different $\eta $ for the Nb isotopes; as it was obtained from the fitting procedure investigating the $^{89}$Y$(\alpha ,3n)^{90}$Nb$^{m,g}$ reaction. 
The result is depicted in Fig.~\ref{fig:eta-effec-cs} which also shows the ratio of the calculated cross sections. 
Using the evaluated $\eta $ values the maximum change was about 300 percent. 
The $\eta $ values do not modify the maximum cross section, but they modify the high energy part of the excitation function. 
We did not want to compare the calculation with the experimental data, but only to demonstrate that this systematics will have an effect on the cross section calculation.

\section{Summary}

We have developed a method for the determination of $\eta =\Theta _{eff}/\Theta _{rig}$, a parameter of the spin distribution of the level density of an excited nucleus, from the measured isomeric cross-section ratios by comparing them with the model calculation using the TALYS 1.6 code. 
The minimum of the reduced $\chi ^2$ of the ratio of the measured to calculated isomeric cross-section was determined as a function of the $\eta $ and the number of used discrete levels~(N). 
The isomeric cross-section ratios were analyzed for 25 final nuclei. 
The nuclei in the chain between the first compound nucleus to the isomeric pair contribute to isomeric cross-section ratio, therefore, in many cases, more than one $\eta $ values were determined. 
In some cases, more than one reactions were used to produce the same isomeric pair which also increased the total number of the evaluated $\eta $ values to 61. 

The $\eta_d $ values were derived, using Eq.~\ref{eq:deta}, for the evaluated nuclei from the low discrete levels. 
The ratio $\eta /\eta_d $ obtained from the continuum and the discrete levels is constant as a function of the mass number, its value is near to 1.0 but this constant shows a small dependence on the even-odd type of the neutron and proton number. 
Using this systematics the $\eta $ value for any nucleus can be estimated based on the $\eta_d $ values. This systematics was successfully applied to extend the $\eta $ data for even-even type nuclei.

Considering the $\eta /\eta_d $ ratio as a function of excitation energies of the nuclei, the $\eta $ values were found to be independent of the excitation energy of the nucleus.

We have analyzed the mass number (A) and (N-Z) dependence of the $\eta $ values. In both nuclei with odd masses show an exponential decrease with slightly different functions for odd(N)-even(Z) and even(N)-odd(Z) type of nuclei. The fitted parameters with their uncertainties are presented. 
In the case of even mass nuclei, the even-even and odd-odd type nuclei show different behavior. 
The $\eta $ values for odd-odd type nuclei are constant up to (A) equal to 110 or (N-Z) equal to 13, and then they follow exponential decrease characteristics like those of the nuclei with the odd mass number. 
The $\eta $ values for even-even type nuclei decrease exponentially up to (A) equal to 142 or (N-Z) equal to 20. The values are lower than  for the nearby odd-odd type nuclei. The $\eta $ values  increase for A between 142 and 156 and (N-Z) between 20 and 28; they finally reach the value of odd-odd type systematics.
The $\eta $ values above A=156 and (N-Z)= 28 follow the exponential decreasing trend of odd-odd type nuclei. We interpreted this trend change by assuming that $^{156}$Gd consists of a very stable core: magic number of 126 nucleons (63 proton-neutron pairs) and a magic number of 28 neutrons, plus one neutron and a proton as the odd-odd type nuclei.

 The calculation on the $^{89}$Y($\alpha$,xn)$^{92,91,90,89}$Nb reactions demonstrated that the even-odd characteristics of $\eta $, on the another hand, has a significant effect on all the calculated activation cross sections, not only on the isomeric cross-sections.
 
 As a side effect, the excitation energy of  $^{182m}Re$ was determined as 550 keV, and the spin of the  $^{122m}Sb$ was set to $6^- $.

\begin{acknowledgments}
We thank the TALYS team for developing a very well organized program with well defined input parameters and clear output structure, and with the possibility to use different $\eta $ values for even nuclei. Unix script files were used to generate the input files and to collect the needed information from the outputs. These were very important for us to be able to handle the many thousands of calculations of the excitation functions performed in this evaluation.
 
\end{acknowledgments}


\bibliography{eta-aw2}

\end{document}